\documentclass[aps,prd,amsmath,amssymb,twocolumn,nofootinbib,showpacs,preprintnumbers,superscriptaddress]{revtex4-1}
\usepackage{graphics,hyperref,color,bbold,multirow}
\usepackage[utf8]{inputenc}
\usepackage{comment}
\usepackage{tikz}
\hypersetup{ 
    pdfnewwindow=true,      
    colorlinks=true,       
    linkcolor=blue,          
    citecolor=blue,        
    filecolor=blue,      
    urlcolor=blue        
}  
\begin{document}

\title{Bottomonium spectrum with a Dirac potential model in the momentum space} 
\author{David~Molina}
\email{djmolinab@unal.edu.co}
\affiliation{Universidad Nacional de Colombia, Bogot\'a 111321, Colombia}
\author{Maurizio De Sanctis}
\affiliation{Universidad Nacional de Colombia, Bogot\'a 111321, Colombia}
\author{C\'esar Fern\'andez-Ram\'{\i}rez}
\email{cesar.fernandez@nucleares.unam.mx}
\affiliation{Instituto de Ciencias Nucleares, Universidad Nacional Aut\'onoma de M\'exico, 
Ciudad de M\'exico 04510, Mexico}
\author{Elena Santopinto}
\email{santopinto@ge.infn.it}
\affiliation{INFN, Sezione di Genova, via Dodecaneso 33, 16146 Genova, Italy}

\begin{abstract}
We study  the bottomonium spectrum using a relativistic potential model in the momentum space.
 This model is based on a complete one gluon exchange interaction with
a momentum dependent screening factor
to account  for the  effects due to virtual pair creation
that  appear  close  to  the  decay  thresholds. 
The overall model does not make use of  nonrelativistic approximations.
We fit well established bottomonium states below the open charm threshold and predict the
rest of the spectrum up to $\approx 11200$ MeV and $J^{PC}=3^{--}$.
Uncertainties are treated rigorously and propagated in full to the parameters of the model
using a Monte Carlo to identify if which
deviations from experimental data can be absorbed into the statistical uncertainties 
of the models and which can be related to physics beyond the $b\bar{b}$ picture, 
guiding future research.
We get a good description of the spectrum, in particular the Belle measurement of the
$\eta_b(2S)$ state and the $\Upsilon(10860)$  and $\chi_b(3P)$ resonances.
\end{abstract}
\date{\today}
\maketitle

\section{Introduction}
The heavy quark meson sector constitutes a major piece of information on the
nonperturbative regime of the strong interaction.
In particular, a lot of experimental information has been gathered on the bottomonium spectrum
during the last years thanks to ATLAS, BaBar, Belle, BESIII, CLEO, CMS, D0, and LHCb
collaborations~\cite{Aad:2011ih,Abazov:2012gh,Park:2017pne,Sirunyan:2019osb,Aaij:2019evc,Fulsom:2018hpf,Sirunyan:2018dff,Tamponi:2018cuf,BESIII:2016adj,Ablikim:2016qzw,Olsen:2017bmm,PhysRevD.98.030001}
and further results are expected in the near future during the 
Belle II run~\cite{Tanida:2019yvl,Pedlar:2018tck} and 
after the CMS and LHCb upgrades~\cite{Bediaga:2018lhg,Piucci:2017kih}.
Theory work has preceded and followed through the experimental 
effort~\cite{Guo:2017jvc,Lebed:2016hpi,Yuan:2018inv,Liu:2019zoy,Brambilla:2019esw} in the form of 
Lattice QCD computations~\cite{
Liao:2001yh,Meinel:2009rd,Meinel:2010pv,Daldrop:2011aa,Lewis:2012ir,Dowdall:2013jqa,Davies:2013dem,Baker:2015xma,Ding:2018uhl,
Larsen:2019zqv,Bicudo:2019ymo,Larsen:2019bwy}, 
Dyson-Schwinger-Bethe-Salpeter equations~\cite{Fischer:2014cfa,Popovici:2014pha,Hilger:2015hka,Negash:2017rqt,Leitao:2017esb,Mojica:2017tvh,Wang:2019tqf},
and potential quark models~\cite{Godfrey:1985xj,
Gonzalez:2009jk,Liu:2011yp,Ferretti:2013vua,Ferretti:2014xqa,Segovia:2014mca,Akbar:2015evy,Deng:2016ktl,Segovia:2016xqb,Weng:2018ebv,Srivastava:2018vxp,Monteiro:2018dnx,Al-Jamel:2019myn,Chen:2019uzm}.

In this paper we develop a relativistic quark model for bottomonia
based on a complete one gluon exchange.
The approach is completely relativistic and does not rely on nonrelativistic approximations. 
In this way the standard spin-orbit, spin-spin, and tensor interactions are automatically included.
We also incorporate a relativistic scalar interaction and
 a momentum dependent screening factor 
to account  for the  effects  due to  virtual  pair  creation
that  appear  close  to  the  decay  thresholds.
All the calculations are performed in the momentum space.
The same model  was successfully applied to reproduce 
the charmonium spectrum in Ref.~\cite{molina:2017iaa} which we refer the reader to 
for technicalities.
We fit the  model to all the known states of each $J^{PC}$ below the $B\bar{B}$ threshold
except for the recently measured $\chi_{b1}(3P)$ and $\chi_{b2}(3P)$
which we prefer to predict in order to gain insight on their nature
and the $\eta_b(2S)$ which we exclude of our fit
owing to the disagreement between CLEO~\cite{Dobbs:2012zn} and Belle~\cite{Mizuk:2012pb} measurements.
We perform a rigorous error estimation that allows us to assess if the inclusion of a new effect 
in the phenomenological model is necessary or not, and we compute the parameter correlations 
which provide insight on
how independent are the different pieces of the model among them.
A full error analysis is mandatory to identify which deviations from experimental data can be absorbed into the statistical uncertainties of the models and which can be related to physics beyond the $b\bar{b}$ picture, guiding future research.

The paper is organized as follows: In Sec.~\ref{rev_mod} we provide the relativistic quark model
and the employed solution method; In Sec.~\ref{sec:determspectr} 
we describe the fitting procedure as well as the 
statistical method used to compute the uncertainties; 
In Sec.~\ref{sec:bottom} we report
the computed bottomonium spectrum up to $J^{PC}=3^{--}$ and 
$\approx 11200$ MeV as well as the comparison
to the available experimental information.
We obtain a very good description of both fitted 
and nonfitted bottomonia and also predict many
unobserved states; 
Sec.~\ref{conclu} summarizes the conclusions.
     
\section{Model and relativistic equation}\label{rev_mod}
\subsection{Hamiltonian model}
We apply to bottomonia the same model developed in~\cite{molina:2017iaa} for charmonia.
The total interaction Hamiltonian model in the momentum space
is given by the sum of vector ($\bar{{\cal H}}^{(v)}$) and scalar ($\bar{{\cal H}}^{(s)}$) interactions
\begin{equation}\label{hint}
\bar {\cal H}_\text{int} (\vec p_b, \vec p_a)= \bar{{\cal H}}^{(v)}(\vec{p}_b, \vec{p}_a)+\bar{{\cal H}}^{(s)}(\vec{p}_b, \vec{p}_a), 
\end{equation}
where $\vec{p}_a$ and $\vec{p}_b$ represent the three-momenta of both quark and antiquark 
in the center of mass of the bottomonium system.
The vector interaction is based on one gluon exchange, which in the Coulomb gauge reads
\begin{equation}\label{eq4}
\begin{split}
\bar{{\cal H}}^{(v)}(\vec{p}_b, \vec{p}_a)=& \:
V^{(v)}(\vec q)\: \left[ J^0_1\: J^0_2 \left(1-\frac{(\Delta E)^2}{Q^2} \right)\right. \\
&-\left. \vec{J}_1\cdot\vec{J}_2\left(1+\frac{(\Delta E)^2}{Q^2} \right)\right] ,
\end{split} 
\end{equation}
where
\begin{equation}\label{fourcurr}
J_i^{\mu}=  J_i^{\mu} (\vec \sigma_i; \vec p_b,\vec p_a)
=\bar{u} (\vec{p}_{i b},\vec{\sigma}_i)  \gamma_i^{\mu}u(\vec{p}_{i a },\vec{\sigma}_i),
\end{equation}
represents the standard four-current of the quarks, $\vec{\sigma}_i$ stands for the Pauli matrices, and $\gamma^{\mu}_i$ are the gamma matrices,  where $i=1,2$ is the particle label. 
We also introduce the quark energy difference
\begin{equation}\label{delte}
\Delta E= E(\vec p_b) -E(\vec p_a),
\end{equation}
and the squared (positive) four momentum transfer
\begin{equation}\label{defq2}
Q^2=\vec{q}^{\: 2}-(\Delta E)^2.
\end{equation}
where $\vec q= \vec{p}_b- \vec{p}_a $ represents the three momentum transfer.
The scalar interaction is defined as
\begin{equation}\label{scalint}
\bar{{\cal H}}^{(s)}(\vec{p}_b, \vec{p}_a)=V^{(s)}(\vec q) \: I_{1} \: I_{2}, 
\end{equation}
where $I_i$ is a scalar vertex.
Finally, the vector and the scalar effective  potentials have the following form:
\begin{subequations}\label{vtotq}
\begin{equation}\label{vector}
V^{(v)}(\vec q)=-\frac{4}{3}\frac{\alpha_{st}}{\vec{q}^{\: 2}}
+\beta_{v}\frac{3b^2-\vec{q}^{\: 2}}{(\vec{q}^{\: 2}+b^2)^3} ,
\end{equation}
\begin{equation}\label{scalar}
V^{(s)}(\vec q)=A +\beta_{s} \frac{3b^2-\vec{q}^{\: 2}}{(\vec{q}^{\: 2}+b^2)^3} .
\end{equation}
\end{subequations}
Equation~\eqref{vector} represents a regularized Cornell potential, where $\alpha_{st}$ is the coupling constant and $\beta_v$ corresponds to the vector confinement strength. Additionally,  Eq.~\eqref{scalar} 
contains a 
phenomenological constant term  $A$ plus a
$\beta_s$ term  which corresponds to the scalar confinement strength.
The constant parameter $b$ has been introduced to avoid the divergence when $|\vec q|\rightarrow 0$. 

As in~\cite{molina:2017iaa} for charmonia, we use two different prescriptions
for the scalar interaction:
\begin{equation}\label{pot_options}
\begin{cases}  
\text{potential I\phantom{I}} \rightarrow \text{model using Eqs.~(\ref{vtotq}) with } \beta_s=0 , \\
\text{potential II} \rightarrow  \text{model using Eqs.~(\ref{vtotq}) with } \beta_s\neq0 .
\end{cases}
\end{equation}
In this way we can check if the two forms of the scalar interaction (with or without the confinement term) have the same effect on the spectrum, as in the case of the charmonium system.  
Besides, in order to take into account the effects of the virtual \cite{Li:2009nr,Ferretti:2012zz} pair creation that appear close to the decays thresholds,   
we include a screening momentum dependent factor. 
Hence, the total Hamiltonian takes the final form
\begin{equation}\label{screen}
\bar {\cal H}_\text{int} (\vec p_b, \vec p_a)  \rightarrow {\cal H}_\text{int} (\vec p_b, \vec p_a)
=  F_s(p_b) \bar {\cal H}_\text{int} (\vec p_b, \vec p_a)   F_s(p_a),
\end{equation}\label{F_factor}
where the factor $F_s(p)$ is defined as
\begin{equation}
F_s(p)=\frac{1+k_s}{k_s+ \exp{ \left( p^2/p^2_s \right)}}.
\end{equation}
In this way, the  model, with potential I and potential II, depends on seven and eight parameters, respectively.

\subsection{Relativistic equation and solution method}
The relativistic equation we use is obtained performing a three dimensional 
reduction of the Bethe-Salpeter equation 
and keeping only the contributions of the positive energy Dirac spinors~\cite{molina:2017iaa}. 
In the center of mass of the $b \bar b$ system, the relativistic integral equation takes the form
\begin{equation}\label{eigen2}
[K(\vec{p_b})+M_0]\Psi(\vec{p}_b)+\int d^3p_a  {\cal H}_\text{int} (\vec{p}_b,\vec{p}_a)\Psi(\vec{p}_a)=M\: \Psi(\vec{p}_b) ,
\end{equation}
where we have introduced the energy
\begin{equation}\label{kine}
 K(\vec p)= 2\sqrt{\vec{p}^{\: 2} + m^2},
\end{equation}
and $M_0$ represents the phenomenological zero point energy of the spectrum, 
$M$ is the resonance mass (i.e. the eigenvalue of the integral equation) 
and $\Psi(\vec{p})$ is the resonance wave function.
The wave function  $\Psi_{n,\{\nu\}}(\vec{p})$ ($\{\nu\}= L,S,J$) can be written as
\begin{eqnarray}\label{basis}
\Psi_{n,\{\nu\}}(\vec{p})=R_{n,L}(p;\bar{p}) \left[ Y_L(\hat{p})\otimes\chi_S\right]_{J,M_J} ,
\end{eqnarray}
where $R_{n,L}(p;\bar{p})$ corresponds to the radial function 
in the momentum space with 
$n$ the principal quantum number,
$\bar p$ the variational parameter (with dimensions of momentum), 
$Y_{L,M_L}(\hat{p})$ are the spherical harmonics, 
and $\chi_{S,M_S}$ is the spin function.
To solve Eq.~\eqref{eigen2} we use the variational method.
As trial functions we use a
combination of a finite subset of 
three dimensional harmonic oscillators.
Hence, we can write the Hamiltonian matrix as
\begin{widetext}
\begin{equation}
M_{\{\nu\},n_b,n_a}= M_0 \delta_{n_b,n_a}+
\int  d^3 p \: \Psi_{n_b, \{\nu\}}^{\dag}(\vec{p}) K(\vec{p}) \Psi_{n_a,\{\nu\}}(\vec{p})
+\int d^3 p_b \: d^3 p_a \Psi_{n_b,\{\nu\}}^{\dag}(\vec{p_b}) 
{\cal H}_\text{int} (\vec{p}_b,\vec{p}_a
)
\Psi_{n_a,\{\nu\}}(\vec{p_a}) .
\label{eigen3}
\end{equation}
\end{widetext}
The eigenvalues and the eigenstates are found through the variational method, diagonalizing
and minimizing the $M_{\{\nu\},n_b,n_a}$ matrix in Eq.~\eqref{eigen3}~\cite{molina:2017iaa,DeSanctis:2010zz}. The angular part is solved analytically and the radial part numerically.  
The details can be found in the Appendix of Ref.~\cite{molina:2017iaa}.

\section{Parameter determination} \label{sec:determspectr}
To determine the values of the parameters, the uncertainties, and the theoretical bottomonium spectrum
we fit the experimental masses given in Table~\ref{tab:table6}, i.e.
all the known states of each $J^{PC}$ below the $B\bar{B}$ threshold
except for the recently measured $\chi_{b1}(3P)$ and $\chi_{b2}(3P)$
which we try to predict in order to gain insight on their nature
and the $\eta_b(2S)$ which we prefer to exclude of our fit
owing to the disagreement between CLEO and Belle masses~\cite{Segovia:2016xqb}.
From CLEO data a mass of $9974.6\pm2.3\pm2.1$~MeV~\cite{Dobbs:2012zn}
was obtained while Belle measures~$9999.0\pm3.5^{+2.8}_{-1.9}$~MeV~\cite{Mizuk:2012pb}.
BABAR reports a range value between $9974$ and 
$10015$~Mev \cite{Lees:2011mx}.
The PDG favors the Belle measurement~\cite{PhysRevD.98.030001}, 
therefore we show this experimental value in Table~\ref{tab:table6} and Figs.~\ref{fig:espec1} and \ref{fig:espec2}. 
To perform the fits and the error analysis we
use the bootstrap 
technique~\cite{EfroTibs93,Kass2014,Fernandez-Ramirez:2015fbq,Landay:2016cjw}
and proceed as follows:

\begin{figure*}[th!]
\begin{center}
\rotatebox{0}{\scalebox{0.8}[0.6]{\includegraphics{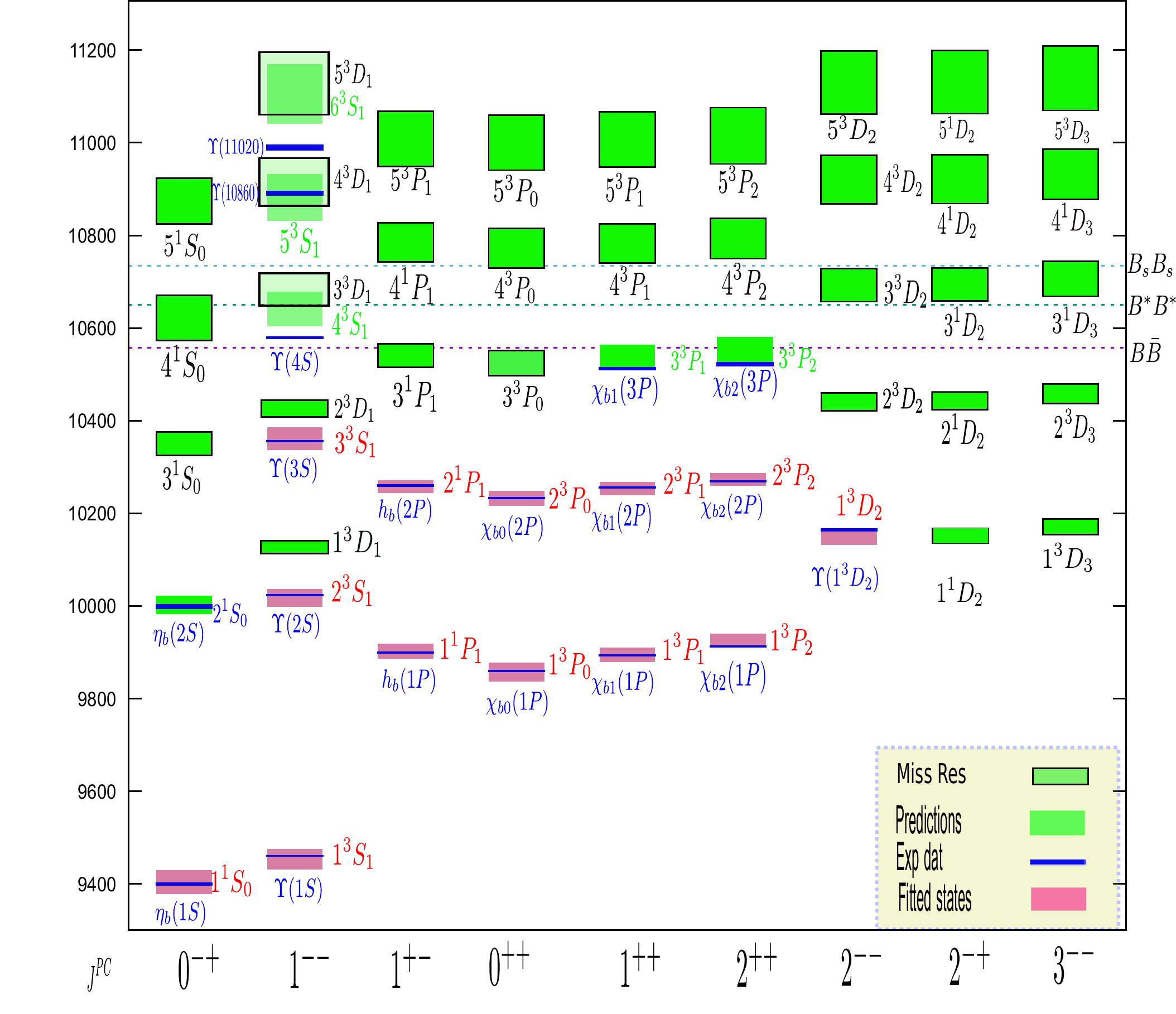}}} 
\caption{Bottomonium spectrum computed with potential I. The blue boxes represent the experimental states
with their error bands, the purple ones provide the computation of the fitted states. The green boxes represent the predictions of the model and, in particular, those with black edges correspond to
missing resonances. For simplicity we only include the names of the experimentally known states. 
}\label{fig:espec1}
\end{center}
\end{figure*}

\begin{figure*}[th!]
\begin{center}
\rotatebox{0}{\scalebox{0.75}[0.6]{\includegraphics{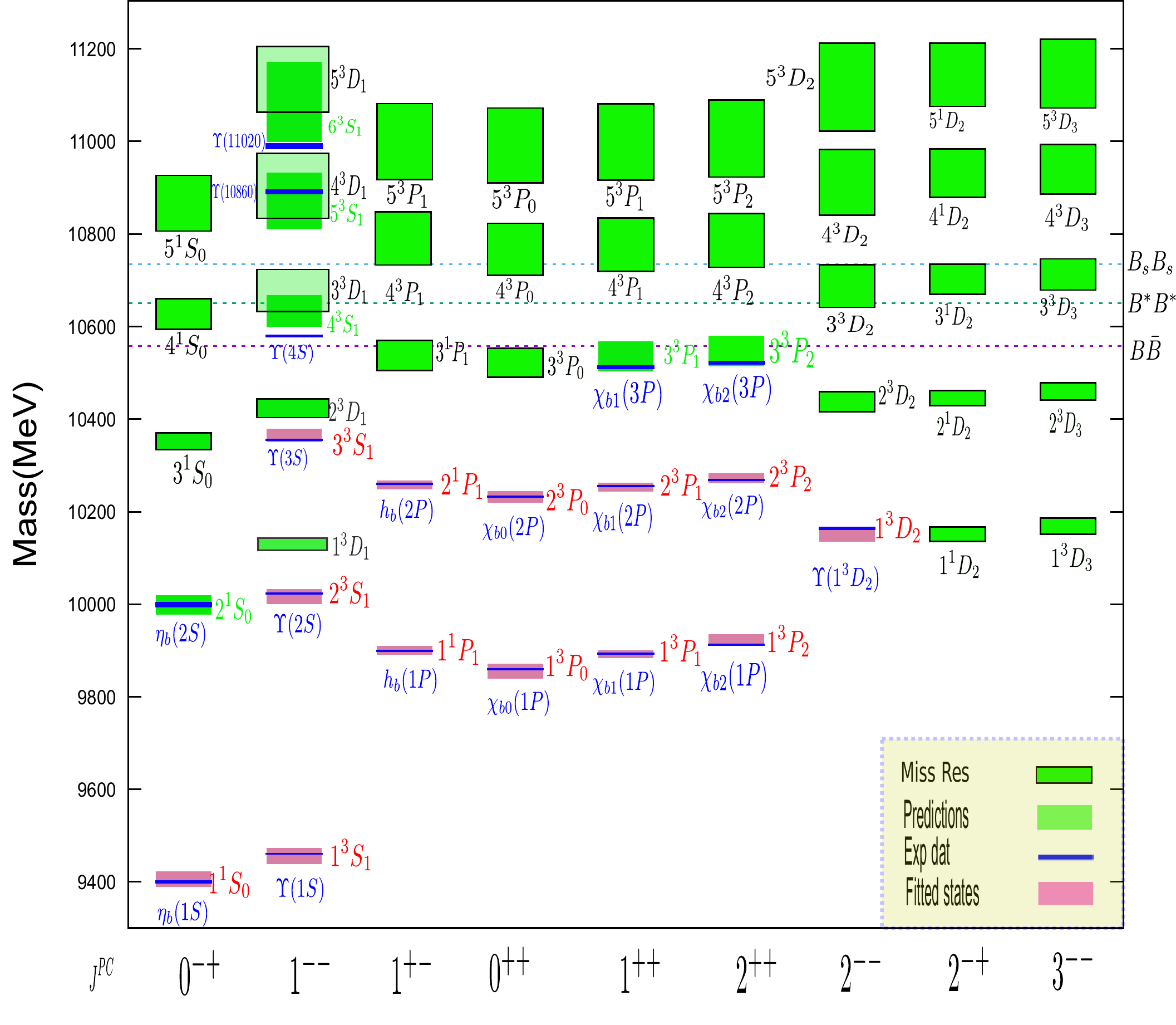}}} 
\caption{Bottomonium spectrum computed with potential II. Same  conventions as in Fig. \ref{fig:espec1}.}\label{fig:espec2}
\end{center}
\end{figure*}

\begin{table*}
\caption{Fitted bottomonia for potentials I and II compared to the PDG values;
$n$ stands for the principal quantum number,
$L$ for the orbital angular momentum, 
$J$ for the total angular momentum,
and $S$ for the spin. The statistical and systematic  errors have been added in quadrature 
for the bootstrap technique.} 
\label{tab:table6}
\begin{ruledtabular}
\begin{tabular}{c|ccccc}
Name &$n\:  ^{2S+1}L_{J}$& \multicolumn{3}{c}{Masses (MeV)} \\ 
& & Potential I  & Potential II& Experiment\\ 
\hline
$\eta_b$      &$1\: ^1S_0$&$\phantom{0} 9402^{+ 27}_{- 24}$& $\phantom{0}9404^{+19}_{-14}$ &$\phantom{0}9399.0 \pm 2.3$  \\
$\Upsilon(1S)$&$1\:^3S_1$&$\phantom{0}9455^{+21}_{-23}$&$\phantom{0}9454^{+19}_{-16}$ &$\phantom{0}9460.30\pm 0.26$\\
$\chi_{b0}(1P)$&$1\:^3P_0$&$\phantom{0}9856^{+ 22}_{- 20}$&$\phantom{0}9858^{+14}_{-19}$
&$\phantom{0}9859.44\pm0.42\pm0.31$\\
$\chi_{b1}(1P)$&$1\:^3P_1$ &$\phantom{0}9894^{+ 17}_{- 15}$&$\phantom{0} 9893^{+9}_{-11} $&$\phantom{0}9892.78\pm0.26\pm0.31$\\
$h_b(1P)$&$1\:^1P_1$&$\phantom{0}9902^{+17}_{-16}$&$9901^{+9}_{-10}$&$\phantom{0}9899.3\pm0.8$\\
$\chi_{b2}(1P)$&$1\:^3P_2$&$\phantom{0}9927^{+ 15}_{- 17}$&$\phantom{0}9923^{+13}_{-14}$&$\phantom{0}9912.21\pm 0.26\pm 0.31$\\
$\Upsilon(2S)$&$2\: ^3S_1$&$\phantom{0}10017^{+ 20}_{- 19}$&$\phantom{0}10016^{+17}_{-15}$&$\phantom{0}10023.26 \pm 0.31$ \\
$\Upsilon(1D)$&$1\:^3D_2$&$\phantom{0}10151^{+13}_{-19}$&$\phantom{0}10149^{+16}_{-14}$&$\phantom{0}10163.7\pm1.4$\\
$\chi_{b0}(2P)$&$2\:^3P_0$&$\phantom{0}10232^{+ 18}_{- 16}$&$\phantom{0}10233^{+12}_{-13}$&$\phantom{0}10232.5\pm 0.4\pm 0.5$\\
$\chi_{b1}(2P)$&$2\:^3P_1$&$\phantom{0}10253^{+14}_{-15}$&$\phantom{0}10254^{+7}_{-11}$&$\phantom{0}10255.46\pm 0.22\pm 0.50$\\
$h_b(2P)$&$2\:^1P_1$&$\phantom{0}10257^{+14}_{-15}$&$\phantom{0}10259^{+8}_{-10}$&$\phantom{0}10259.8\pm0.5\pm 1.1$\\
$\chi_{b2}(2P)$&$2\:^3P_2$&$\phantom{0}10274^{+13}_{-15}$&$\phantom{0}10274^{+11}_{-12}$&$\phantom{0}10268.65\pm 0.22\pm 0.50$\\
$\Upsilon(3S)$&$3\:^3S_1$&$\phantom{0}10361\pm25$&$\phantom{0}10364\pm14$&$\phantom{0}10355.2\pm 0.5$
\end{tabular}
\end{ruledtabular}
\end{table*}

\begin{enumerate}
    \item We randomly choose values for the masses of the resonances by sampling a Gaussian distribution according to their uncertainties 
    (systematic and statistical summed in quadrature), 
    obtaining a resampled bottomonium spectrum; 
    \item We use the least-squares method to minimize the squared distance
    \begin{equation} 
d^2 = \sum_{i} \left( E^{th}_i - M_i \right)^2,
\end{equation} 
where $M_i$ are the resampled experimental bottomonia, i.e.
the states $0^{-+}$ ($\eta_b(1S)$); 
 $1^{--}$ ($\Upsilon(1S)$, $\Upsilon(2S)$,  $\Upsilon(3S)$); 
$0^{++}$ ($\chi_{b0}(1S)$, $\chi_{b0}(2S)$);
$1^{+-}$ ($h_b(1P)$, $h_b(2P)$); 
$1^{++}$ ($\chi_{b1}(1P)$, $\chi_{b2}(2S)$ );
$2^{--}$ ($\Upsilon(1D)$) y
$2^{++}$ ($\chi_{b2}(1P)$, $\chi_{b2}(2S)$). 
The $E^{th}_i$ represents the theoretical states calculated by solving
the eigenvalue Eq.~\eqref{eigen3} with potentials I and II. 
The fit is performed using  MINUIT~\cite{James:1975dr}.
\end{enumerate}

\begin{table}[h]
\caption{Fit parameters for both potentials. 
Error bars are reported at 1$\sigma$ (68\%) CL 
and take into account all the correlations among the parameters.} 
 \label{tab:table1}
\begin{ruledtabular}
\begin{tabular}{c|cc}
Parameter&Potential I&Potential II\\\hline
$m$ (GeV)  &$ 4.52^{+ 0.13}_{- 0.13}$     &$4.51^{+0.08}_{-0.09}$\\
$M_0$ (GeV) &$\phantom{0}0.48^{+ 0.33}_{-0.27}$       &$\phantom{0}0.47\pm0.2$\\
$\alpha_{st}$ &$\phantom{00}0.39^{+  0.09}_{-0.10}$ &$\phantom{0}0.37^{+ 0.10}_{-0.10}$\\
$\beta_{v}$ (GeV$^2$)&$\phantom{0}0.018^{+0.004}_{-0.001}$ &$\phantom{0}0.017^{+ 0.003}_{-0.003}$\\
$k_{s}$  &$\phantom{00}98^{+22}_{-12}$ &$\phantom{0}100^{+29}_{-20}$\\
$p_{s}$ (GeV) &$\phantom{0}1.55^{+ 0.23}_{-0.20}$ &$\phantom{0}1.56^{+ 0.23}_{-0.21}$\\
$A$ (GeV$^{-2}$) &$\phantom{00}0.0011\pm 0.0010$  &$-0.0013\pm0.0013$\\
$\beta_{s}$ (GeV$^2$)&$\phantom{00}0$ (fixed) &$\phantom{0}0.090^{+ 0.002}_{-0.002}$
\end{tabular}
\end{ruledtabular}
\end{table}

This procedure is repeated 1000 times in order to obtain enough statistics 
to compute the expected values of the parameters as well as their uncertainties at a $1\sigma$ ($68\%$) confidence level (CL).
The expected value of the parameters (Table~\ref{tab:table1})
are computed as the mean value of the 1000 samples.
The uncertainties are obtained as the 
the differences between the mean value and the highest and lowest masses of the 
best 68\% of the fits. Hence, our uncertainties can be asymmetric.
Once the parameters have been determined, we can compute the 
bottomonium spectrum and the associated uncertainties
(Table~\ref{tab:table6}). 
We find an excellent agreement between theory and fitted states within
uncertainties.
We  note that the values of the common parameters of the two potentials are very similar. These results show that, unlike  for charmonia, the scalar
confinement term of the interaction does not seem to be relevant in the bottomonia description. 
To gain further insight on this issue we compute the
correlation matrices, Tables~\ref{correlation1} and~\ref{correlation2},
for potentials I and II, respectively.

\begin{table}
\caption{Correlation matrix for the parameters of potential I.} \label{correlation1}
\begin{ruledtabular}
\begin{tabular}{c|ccccccc}
&$m$         &$M_0$             &$\alpha_{st}$&$\beta_v$&$k_{s}$&$p_{s}$&$A$ \\\hline
$m$          &  $\phantom{-}1$  & & &  & &\\
$M_0$        &  $-0.89$         & $\phantom{-}1$   & && &\\
$\alpha_{st}$& $\phantom{-}0.13$& $\phantom{-}0.30$&$\phantom{-}1$&& &\\
$\beta_v$    & $-0.08$          & $-0.36$          &$       -0.87$&$\phantom{-}1$& & &\\
$k_{s}$      & $-0.03$          &$\phantom{-}0.01$ & $      -0.09$&$\phantom{-}0.03$&$\phantom{-}1$  &  &\\
$p_{s}$      & $\phantom{-}0.09$&$          -0.09$ & $-0.07$&$\phantom{-}0.08$&$\phantom{-}0.02$&$\phantom{-}1$&\\ 
$A$          & $-0.18$          &$          -0.10$ &$      -0.68$ &$\phantom{-}0.46$&$-0.12$&$-0.55$&$\phantom{-}1$ \\
\end{tabular}
\end{ruledtabular}
\end{table}

\begin{table}
\caption{Correlation matrix for the parameters of potential II.} \label{correlation2}
\begin{ruledtabular}
\begin{tabular}{c|cccccccc}
&$m$&$M_0$&$\alpha_{st}$&$\beta_v$&$k_{s}$&$p_{s}$&$A$ &$\beta_s$\\ \hline
$m$  &$\phantom{-}1$ & & && && &\\
$M_0$ &$-0.76$  &  $\phantom{-}1$ & & && & &\\
$\alpha_{st}$&$\phantom{-}0.53$ &$\phantom{-}0.14$ &$\phantom{-}1$ & & & & &\\
$\beta_v$&$-0.26$  &$-0.15$   &$-0.58$     &$\phantom{-}1$ & & & &\\
$k_{s}$ & $\phantom{-}0.34$  &$-0.29$  &$\phantom{-}0.06$ &$-0.04$ &$\phantom{-}1$ & & &\\
$p_{s}$ &$\phantom{-}0.13$  &$-0.13$  &$-0.05$   &$\phantom{-}0.03$ &$\phantom{-}0.36$ &$\phantom{-}1$ & &\\
$A$ & $\phantom{-}0.42$  &$-0.05$  &$\phantom{-}0.52$ &$-0.32$ &$\phantom{-}0.48$&$\phantom{-}0.77$ &$\phantom{-}1$ &\\
$\beta_s$& $-0.21$ &$\phantom{-}0.01$  &$-0.31$ &$-0.57$&$-0.03$ &$\phantom{-}0.07$&$-0.07$ &$\phantom{-}1$\\
\end{tabular}
\end{ruledtabular}
\end{table}

For potential I (Table~\ref{correlation1}) 
we find a strong correlation between the parameters of the vector
interaction ($\alpha_s$ and $\beta_v$) and
the scalar interaction parameter $A$,
which indicates  
that vector and scalar interactions 
are physically correlated in this model.
The screening parameter $p_s$ is weakly correlated  
with the vector interaction parameters 
but strongly correlated with the scalar interaction ones.  
For potential II,  we have the additional parameter$\beta_s$.
In this potential, the parameters are less correlated
as shown in  Table~\ref{correlation2} with one exception,
the additional scalar interaction parameter $\beta_s$
is noticeble correlated with the vector interaction parameter $\beta_v$.
Consequently, we find a significant correlation
between the confinement terms 
of the vector and the scalar interactions.  
The parameter $p_s$ of the screening factors is weakly correlated 
with the other parameters of the interactions except with 
the phenomenological parameter $A$ in the scalar interaction.  
This sizeable 
correlation highlights how the screening 
factor impacts more on the scalar interaction.

\begin{figure}
\begin{center}
\rotatebox{0}{\scalebox{0.38}[0.3]{\includegraphics{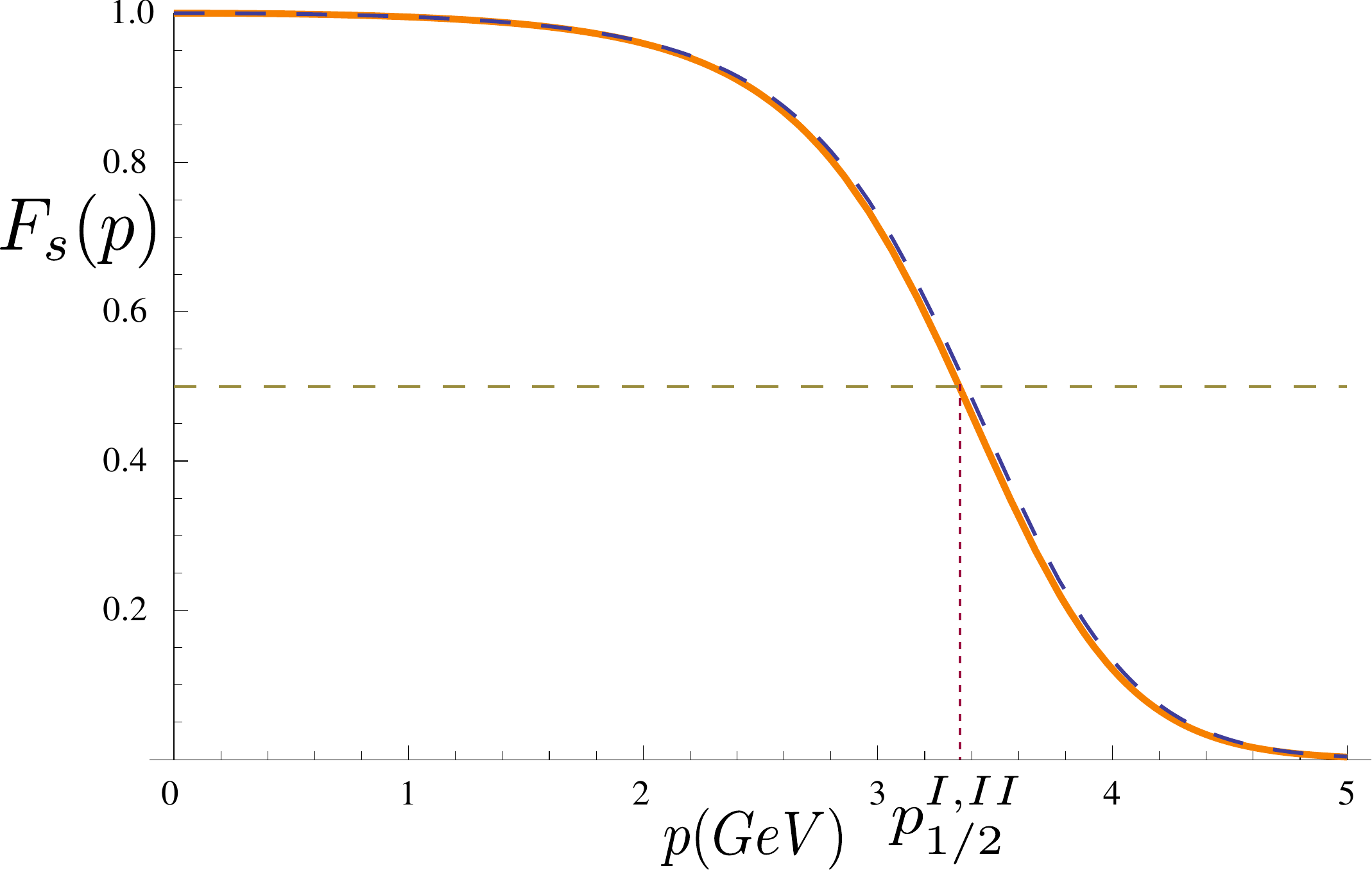}}} 
\caption{Screening function for potentials I (dashed blue) and II (solid orange).
For $k_s$ and $p_s$  we use the central values in Table \ref{tab:table1}.
We highlight the values of the screening momenta 
$p^\text{I}_{1/2}$ (potential I) and $p^\text{II}_{1/2}$ (potential II)
introduced in \cite{molina:2017iaa}.}\label{Screening}
\end{center}
\end{figure}

Using the values obtained in the fitting procedure we plot the screening function $F_s(p)$ in Fig.~\ref{Screening} 
for the two potentials.
As  mentioned above we introduce the screening momenta $p^{\: \text{j}}_{1/2}$ ($\text{j}=\text{I}, \text{II}$ 
labels potentials I and II) which are given 
by $F_s(p^{\: \text{j}}_{1/2})=1/2$ (we recall that $F_s(0)=1$).
Through the fitting values, we find
$p^\text{I}_{1/2}= 3.38~\text{GeV}$ and  $p^\text{II}_{1/2}= 3.34~\text{GeV}$.
These values correspond to the screening kinetic energy
\begin{equation}
\bar{E}^{\: \text{j}}=2\sqrt{m^2 + \left( p^{\: \text{j}}_{1/2}\right) ^{\: 2} },
\end{equation}
which amount to
$\bar{E}^\text{I}=11.281~\text{GeV}$ for potential I and $\bar{E}^\text{II}=11.260~\text{GeV}$ for potential II.
This result show that the screening effect is active above the open bottom threshold as in charmonia. Nevertheless, due to the high values of $\bar{E}^\text{I,II}$, 
we find that the screening effect is less relevant for the low-lying part of the bottomonium spectrum than 
for charmonia~\cite{molina:2017iaa}.
 
\section{Bottomonium spectrum}\label{sec:bottom}
Using the relativistic model interaction, 
with either potential I or II,
we obtain the bottomonium spectrum.
Through the bootstrap method, 
the errors in the fitted states are carried in full to the 
computed uncertainties in the parameters and to the 
spectrum.
We provide the computed spectrum in
Tables~\ref{tab:table6} (fitted states) 
and~\ref{tab:table7} (predicted states). 
The computed and the experimental spectra are compared in 
Figs.~\ref{fig:espec1} (potential I) and~\ref{fig:espec2} 
(potential II).  
In general, the spectrum is reproduced by the model 
within the experimental uncertainties.

We note that the parameters obtained with both potentials are very similar, leading to closely akin spectra.
This result shows that the confining part of the scalar potential does not impact
the bottomonium spectrum. 
However, the presence of the scalar interaction
is necessary for an optimum fit, 
i.e. the parameter $A$ 
contribution in Eq.~\eqref{scalar}.
In what follows we look into the states that were not included in the fit as well as the predicted higher-lying spectrum.

\begin{table*}
\caption{
Predicted bottomonia 
for  potentials I and II compared to the existing experimental masses with their corresponding uncertainties. 
Notation as in Table~\ref{tab:table6}. 
}\label{tab:table7}
 \begin{ruledtabular}

\begin{tabular}{c|cccc}
\multicolumn{1}{c|}{Name}            & $n\:  ^{2S+1}L_{J}$ & \multicolumn{3}{c}{Mass (MeV)}                                                                                                          \\
\multicolumn{1}{c|}{}                  &                     & Potential I                            & Potential II                                               & Experiment                        \\ \cline{1-5} 
\multicolumn{1}{c|}{$\eta_b(2S)$}      & $2\:^1S_0$          & $10000^{+21}_{-17}$    & $9999^{+20}_{-22}$                              & $9999.0\pm 3.5^{+2.8}_{-1.9}$\cite{Mizuk:2012pb}         \\
\multicolumn{1}{c|}{$---$}             & $1\:  ^1D_2$        & $10153^{+14}_{-69}$         & $10150^{+ 16}_{- 14}$                              & $---$                              \\
\multicolumn{1}{c|}{$---$}               & $1\: ^3D_1$         & $10130^{+ 11}_{- 17}$ & $10128^{+14}_{-12}$                     & $---$                              \\
\multicolumn{1}{c|}{$---$}             & $1\:  ^3D_3$        & $10173^{+15}_{-19}$         & $\phantom{0.}10169\pm 17$                                  & $---$                              \\
\multicolumn{1}{c|}{$---$}               & $2\:^1D_2$         & $10445^{+17}_{-21}$         & $10446^{+ 15}_{- 16}$       & $---$                              \\
\multicolumn{1}{c|}{$---$}               & $2\:^3D_1$         & $10427^{+17}_{-19}$   & $10429^{+ 14}_{- 26}$       & $---$                              \\
\multicolumn{1}{c|}{$---$}               & $2\:^3D_2$         & $10443^{+17}_{-21}$         & $10444^{+ 15}_{- 28}$ & $---$                              \\
\multicolumn{1}{c|}{$---$}              & $2\:^3D_3$         & $10460^{+19}_{-23}$         & \multicolumn{1}{c}{$10461^{+18}_{-20}$}       & $---$                              \\
\multicolumn{1}{c|}{$\eta_b(3S)$}             & $3\:^1S_0$          & $10351^{+26}_{-25}$   & \multicolumn{1}{l}{\phantom{0}$10353^{+ 16}_{- 19}$} & $---$                              \\
\multicolumn{1}{c|}{$h_b(3P)$}             & $3\:^1P_1$          & $10542^{+24}_{-26}$         & \multicolumn{1}{l}{\phantom{0}$10546^{+ 23}_{- 40}$}       & $---$                              \\
\multicolumn{1}{c|}{$\chi_{b0}(3P)$}            & $3\:^3P_0$          & $10523^{+28}_{-26}$   & \multicolumn{1}{l}{\phantom{0}$10528^{+ 25}_{- 38}$}       & $---$                              \\
\multirow{2}{*}{$\chi_{b1}(3P)$}      & \multirow{2}{*}{$3\:  ^3P_1$} & \multirow{2}{*}{$10538^{+26}_{-27}$}        & \multirow{2}{*}{$10541^{+24}_{-41}$}                          & \multirow{2}{4.0cm} {$10515.7^{+2.2}_{-3.9}\phantom{:}^{+1.5}_{-2.1}$~\cite{Aaij:2014hla} $10513.42\pm 0.41 \pm  0.18$~\cite{Sirunyan:2018dff}}       \\
  &
  &
  &
  &
  \\
\multicolumn{1}{c|}{$\chi_{b2}(3P)$}                    & $3\:^3P_2$          & $10554^{+25}_{-28}$         & \multicolumn{1}{l}{\phantom{0}$10557^{+ 22}_{- 42}$}       & $10 524.02 \pm 0.57  \pm  0.18$~\cite{Sirunyan:2018dff}                              \\
\multicolumn{1}{c|}{$---$}    & $3\:^3D_2$          & $10697^{+ 33}_{-39 }$       & $10701^{+ 59}_{- 32}$                             & $---$         \\
\multicolumn{1}{c|}{$---$}    & $3\:^1D_2$          & $10699^{+32 }_{-39 }$       & $10702^{+ 32}_{-32}$                             & $---$         \\
\multicolumn{1}{c|}{$---$}    & $3\:^3D_3$          & $10711^{+34 }_{- 41}$       & $10714^{+ 35}_{- 32}$                             & $---$         \\
\multicolumn{1}{c|}{$---$}    & $3\:^3D_1$          & $10685^{+ 31}_{- 37}$       & $10689^{+ 33}_{- 29}$                             & $---$         \\
\multicolumn{1}{c|}{$\eta_b(4S)$}              & $4\:^1S_0$ & $10635^{+ 37}_{- 39}$       & $10638^{+22}_{-44}$                 & $---$         \\
\multicolumn{1}{c|}{$h_b(4P)$}          & $4\:^1P_1$ & $10787^{+41}_{-43 }$       & $10792^{+ 43}_{-71 }$                             & $---$         \\
\multicolumn{1}{c|}{$\chi_{b0}(4P)$}    & $4\:^3P_0$ & $10773^{+ 42}_{-44 }$       & $10779^{+ 43}_{- 69}$                             & $---$         \\
\multicolumn{1}{c|}{$\chi_{b1}(4P)$}    & $4\:^3P_1$ & $ 10785^{+43}_{-42 }$       & $10790^{+ 44}_{- 71 }$                             & $---$         \\
\multicolumn{1}{c|}{$\chi_{b2}(4P)$}    & $4\:^3P_2$ & $10796^{+42 }_{-45}$       & $10801^{+ 43}_{- 72 }$                             & $---$         \\
\multicolumn{1}{c|}{$---$}              & $4\:^3D_2$ & $10926^{+49 }_{-56 }$       & $10929^{+ 53}_{-89 }$                             & $---$         \\
\multicolumn{1}{c|}{$---$}              & $4\:^1D_2$ & $10927^{+49 }_{-56 }$       & $10930^{+ 53}_{- 51}$                             & $---$         \\
\multicolumn{1}{c|}{$---$}              & $4\:^3D_3$ & $10937^{+ 51}_{-58 }$       & $10940^{+ 53}_{-54 }$                             & $---$         \\
\multicolumn{1}{c|}{$---$}              & $4\:^3D_1$ & $10915^{+ 48}_{- 54}$       & $10920^{+ 52}_{- 48}$                             & $---$         \\
\multicolumn{1}{c|}{$\Upsilon(4S)$}     & $4\:^3S_1$          & $10642^{+ 36}_{- 39}$       & $10646^{+21}_{-46}$                             & $10579.4\pm1.2$ \cite{PhysRevD.98.030001}        \\
$\eta_b(5P)$                              & $5\:^1S_0$ &$10878^{+47 }_{-51 }$        & $10883^{+ 42}_{-78 }$                               & $---$ \\
$h_b(5P)$                               & $5\:^1P_1$ &$11013^{+ 58}_{-61 }$        & $11018^{+ 62}_{-101 }$                               & $---$ \\
$\chi_{b0}(5P)$                         & $5\:^3P_0$ &$11002^{+ 60}_{-59 }$        & $11008^{+ 62}_{-99}$                               & $---$ \\
$\chi_{b1}(5P)$                         & $5\:^3P_1$ &$11011^{+58 }_{-61 }$        & $11017^{+  63}_{- 101}$                               & $---$ \\
$\chi_{b2}(5P)$                         & $5\:^1P_2$ &$11020^{+59 }_{-63 }$        & $11025^{+ 63}_{- 103}$                               & $---$ \\
$---$                                   & $5\:^3D_2$ &$11137^{+65 }_{-72 }$        & $11137^{+ 73}_{- 116}$                               & $---$ \\
$---$                                   & $5\:^1D_2$ &$11138^{+64 }_{-72 }$        & $11138^{+ 73}_{-63 }$                               & $---$ \\
$---$                                   & $5\:^3D_3$ &$11146^{+66 }_{-74 }$        & $11146^{+74 }_{- 74}$                               & $---$ \\
$---$                                   & $5\:^3D_1$ &$11128^{+ 64}_{- 70}$        & $11129^{+ 72}_{- 69}$                               & $---$ \\
\multicolumn{1}{c|}{$\Upsilon(10860)$} & $5\:^3S_1$          & $10884^{+ 48}_{- 53}$       & $10889^{+ 43}_{- 79}$                           & $10889.9^{+3.2}_{-2.6}$   \cite{PhysRevD.98.030001}     \\
\multicolumn{1}{c|}{$\Upsilon(11020)$} & $6\:^3S_1$          & $11107^{+ 62}_{- 66}$       & $11108^{+ 64}_{-107}$                           & $10992.9^{+10.0}_{-3.1}$ \cite{PhysRevD.98.030001}
\end{tabular}
\end{ruledtabular}
\end{table*}

\subsection{$\Upsilon(4S)$, $\Upsilon(10860)$ and $\Upsilon(11020)$} 
These resonances belong to the family with quantum numbers $1^{--}$. They were discovered  by means of  $e^{+}e^{-}$ collisions in the mid-eighties~\cite{Besson:1984bd,Lovelock:1985nb}
and were more recently measured by the Belle collaboration~\cite{Santel:2015qga}. 
The $\Upsilon(4S)$ is regarded as a $4^3S_1$ state; its experimental mass is $M_{\Upsilon(4S)}=10579.4\pm 1.2$ MeV
and is not well reproduced by either potential I or II.
This resonance is generally considered as a $b\bar{b}$ state,  but its mass is overestimated by models that make use of different approaches: e.g., the nonrelativistic model in Ref.~\cite{Godfrey:1985xj} provides $M_{\Upsilon(4S)}\simeq 10630$ MeV, the semirelativistic model of Ref.~\cite{Segovia:2016xqb} finds $M_{\Upsilon(4S)}=10607$ MeV, and the non-relativistic coupled channels model in Ref.~\cite{Liu:2011yp} reports $M_{\Upsilon(4S)}=10603$ MeV.
Our computations provides approximately $10642\pm40$ MeV, with both potentials. This result is
compatible with the other models, but far away from the experimental value, even when the uncertainties are taken into account.
Consequently, our result combined with non-relativistic calculations suggest that 
there must be beyond the $q\bar{q}$ picture effects that need to be included to properly describe the state.

The  $\Upsilon(10860)$ resonance is generally interpreted as a $\Upsilon(5^3S_1)$, e.g. in~\cite{Segovia:2016xqb,Liu:2011yp,Godfrey:1985xj,Weng:2018ebv,Akbar:2015evy}. 
However, the theoretical calculations for the pion emission decay widths, 
to $\Upsilon(1S)$, $\Upsilon(2S)$ and $\Upsilon(3S)$ are two orders of magnitude~\cite{Segovia:2014mca} greater than the measurement~\cite{Abe:2007tk} leading to different possible interpretations, such as that $\Upsilon(10860)$ is 
a mixing of a standard  $\Upsilon(5S)$ with a $P$ hybrid state~\cite{Bruschini:2018lse},
Finally, in Ref.~\cite{Gonzalez:2009jk} this state is interpreted as a  $\Upsilon(6S)$, and, hence, 
the $\Upsilon(5S)$ becomes a missing resonance of the experimental spectrum
In our model, this mass state can be reproduced as a $\Upsilon(5S)$ ($5^3S_1$)
(see Table~\ref{tab:table7} and Figs.~\ref{fig:espec1} and~\ref{fig:espec2}) 
or as a $4^3D_1$ state with both potentials.
We do not find support the $\Upsilon(6S)$ interpretation.
Actually, our predicted mean value mass, with $\Upsilon(5S)$, is only $5$ ($1$) MeV
away from the experimental value with potential I (II). 
Consequently, we identify this state as a  $b\bar{b}$ 
with $\Upsilon(5S)$ quantum numbers.
However, any final conclusion requires the explanation of the 
before mentioned pion emission decay widths which we leave for a future work.

Finally, the $\Upsilon(11020)$ state is mostly described as a $b \bar{b}$ meson in a  $6^3S_1$ state
except for~\cite{Gonzalez:2009jk} which interprets it as a $7^3S_1$ state.
We do not find a satisfactory description of the mass of this state, neither as $6^3S_1$ nor $7^3S_1$ with either potential.
In fact, our results are similar to that of a  nonrelativistic model in~\cite{Godfrey:1985xj}. 
Hence, we favor the existence of additional physics to explain the mass of this state,
such as coupled channel effects
as shown in~\cite{Liu:2011yp} where a mass of $11023$ MeV is obtained, 
very close to the experimental value.

\begin{table*}
\centering
\caption{Theoretical results obtained, using Potential I, for the states  $\chi_{bJ}(nP)$ compared with the available  experimental data; $n=1,2,3$ is the principal quantum number;  $\bar{M}_n$ stands for the barycenter of the system for each $n$. 
The experimental states for $n=1,2$ are taken from Ref.~\cite{Sirunyan:2018dff}.  
The statistical and systematic errors of the experimental states have been summed in quadrature in order to obtain the errors of the experimental barycenter masses. The  theoretical uncertainties of the  barycenters were propagated from the parameters through the bootstrap technique.}
\label{tab:table_chi_model1}
\begin{ruledtabular}
\begin{tabular}{c|cccccc}
\multicolumn{4}{c|}{Theory: Potential I}                     & \multicolumn{3}{c}{Experimental}                                                       \\ \hline
$n$                  & 1        & 2         & 3         & 1                            & 2                              & 3                       \\ \hline
$M_{\chi_{b0}(nP)}(MeV)$ & $9856^{+22}_{-20}$   & $10232^{+18}_{-16}$   & $10523^{+28}_{-26}$   & $\phantom{0}9859.44\pm0.42\pm0.31$  & $\phantom{0}10232.5\pm 0.4\pm0.5$    &  $---$                       \\ 
\multirow{2}{*} {$M_{\chi_{b1}(nP)}(MeV)$} & \multirow{2}{*} {$9894^{+17}_{-15}$} & \multirow{2}{*} {$10253^{+14}_{-15}$}   & \multirow{2}{*} {$10538^{+26}_{-27}$}   & \multirow{2}{*} {$\phantom{0}9892.78\pm0.26\pm0.31$}  & \multirow{2}{*} {$\phantom{0}10255.46\pm0.22\pm0.50$}     & \multirow{2}{4.2cm} {\centering $10512.1\pm 2.1 \pm 0.9 $~\cite{PhysRevD.98.030001} $10513.42\pm 0.41 \pm  0.18$~\cite{Sirunyan:2018dff}} \\ &&&&\\
$M_{\chi_{b2}(nP)}(MeV)$ & $9927^{+15}_{-17}$ & $10274^{+13}_{-15}$ & $10554^{+25}_{-28}$ & $\phantom{0}9912.21\pm 0.26\pm0.31$ & $\phantom{0}10268.65\pm 0.22\pm0.50$  &   $10 524.02 \pm 0.57  \pm  0.18$ \cite{Sirunyan:2018dff}                      \\ \hline
\multirow{2}{*} {\centering$\bar{M}_{nP}(MeV)$ }      &\multirow{2}{*} {\centering $9908\pm15$ }& \multirow{2}{*} {\centering$10262^{+14}_{-15}$} &\multirow{2}{*} {\centering $10545^{+24}_{-27}$} &\multirow{2}{*} {\centering $\phantom{0}9899.87 \pm 0.27$} & \multirow{2}{*} {\centering $\phantom{0}10260.20 \pm 0.36$} &  \multirow{2}{3.2cm}{\centering $10530 \pm 5\pm 9$~\cite{Aad:2011ih} $10551\pm 14\pm17$~\cite{Abazov:2012gh}}\\
&&&&\\
\end{tabular}
\end{ruledtabular}
\end{table*}

\subsection{$\chi_b(3P)$ states}\label{sec_3P}
The $\chi_b(3P)$ states have been the focus of several experimental collaborations during the
last years.
An estimation of the $\chi_b(3P)$ system \textit{barycenter}
(i.e.  spin-weighted mass average of 
the $\chi_{b0}(3P)$, $\chi_{b1}(3P)$, and $\chi_{b2}(3P)$ states)
was reported by ATLAS~ \citep{Aad:2011ih} 
and D0~\cite{Abazov:2012gh}  collaborations,
yielding $10530\pm5(stat)\pm 9(syst)$ MeV and $10551  \pm 14(stat) \pm  17(syst)$ MeV, respectively.
More recently, two out of the three state masses were measured;
$\chi_{b1}(3P)$ by the LHCb collaboration obtaining
$10515.7^{+2.2}_{-3.9} (stat)^{+1.5}_{-2.1} (syst)$ MeV,
and 
$\chi_{b1}(3P)$
and $\chi_{b2}(3P)$  by the CMS collaboration~\cite{Sirunyan:2018dff} yielding 
$10513.42 \pm 0.41 (stat) \pm  0.18 (syst)$ MeV 
and $10524.02 \pm 0.57 (stat) \pm  0.18 (syst)$ MeV, respectively.
Several predictions  of these states are available in the
literature, employing different frameworks.
For example, in Ref.~\cite{Li:2009nr} a mass of  $10524$ MeV
is predicted for the $\chi_{b1}(3P)$ state employing a screened potential;
in Ref.~ \citep{Liu:2011yp}, $10517$ MeV for the same state by means of a
coupled channel calculation;
and $10580$ MeV in the unquenched quark model~\cite{Ferretti:2013vua}.
All of the results overestimate the mass of $\chi_{b1}(3P)$.
In our calculation (which porpously does not fit this state) 
we obtain $\simeq10540\pm 30$ MeV with both potentials
whose central value also overestimates
the mass of the state. 
When the uncertainties are taken into account, the experimental
mass falls within our error bars and no indication of the need for additional 
physics is called for.
This shows how important it is to perform a rigorous error estimation when 
performing a level-by-level comparison between theory and experiment, 
as differences that can be accounted by the error analysis can be mistaken by
physics beyond the $b\bar{b}$ picture.
Regarding $\chi_{b2}(3P)$, 
$10532.4$ Mev is obtained in Ref.~\cite{Liu:2011yp} using the coupled channels formalism
and  $10578$ Mev under the unquenched quark model~\cite{Ferretti:2013vua}.
We obtain $10554^{+25}_{-28}$ and $10557^{+ 22}_{- 42}$ 
with potentials I and II, respectively.
The CMS value falls 
well within our uncertainites for  potential I
and slightly out of them for potential II, although certainly within $2\sigma$ uncertainties.
Hence, the individually measured $\chi_b(3P)$ states are well reproduced by our model.
Finally, we obtain the barycenter mass  
$10545^{+24}_{-27}$ MeV for potential I and 
$10549^{+23}_{-41}$ MeV for potential II,
both compatible with the previously quoted ATLAS and D0 estimations.
Recalling that not all the individual states of the $\chi_{bJ}(3P)$ system 
have been measured,
we provide in Tables~\ref{tab:table_chi_model1} (potential I)
and~\ref{tab:table_chi_model2} (potential II)
the $n=1, 2, 3$ barycenter masses, given by~\cite{Dib:2012vw,Anwar:2018yqm}
\begin{equation}\label{barycenter}
\bar{M}_{nP}=\frac{M_{\chi_{b0}(nP)}+3M_{\chi_{b1}(nP)}+5M_{\chi_{b2}(nP)}}{9},
\end{equation} 
along with 
the available experimental measurements and estimates from PDG values
Given that  both potentials produce similar spectra, the $\chi_{b}(nP)$ barycenters are very similar. 
In summary, we find a good agreement, within errors, between the models and the experimental barycenters.

\begin{table*}
\centering
\caption{Same as in Table~\ref{tab:table_chi_model1} with the results given by Potential II.}

\label{tab:table_chi_model2}
\begin{ruledtabular}
\begin{tabular}{c|cccccc}
\multicolumn{4}{c|}{Theory: Potential II} & \multicolumn{3}{c}{Experimental}                               \\ \hline
$n$                   & 1         & 2          & 3           & 1                            & 2                              & 3                       \\ \hline
$M_{\chi_{b0}(nP)}(MeV)$  & $9858^{+14}_{-19}$    & $10233^{+12}_{-13}$    & $10528^{+25}_{-38}$  & $\phantom{0}9859.44\pm0.42\pm0.31$  & $\phantom{0}10232.5\pm 0.4\pm0.5$    & $---$                   \\     
\multirow{2}{*} {$M_{\chi_{b1}(nP)}(MeV)$}  & \multirow{2}{*} {$9893^{+9}_{-11}$}    & \multirow{2}{*} {$10254^{+7}_{-10}$}  & \multirow{2}{*} {$10541^{+24}_{-41}$}   & \multirow{2}{*} {$\phantom{0}9892.78\pm0.26\pm0.31$}  & \multirow{2}{*} {$\phantom{0}10255.46\pm0.22\pm0.50$}     & \multirow{2}{4.2cm} {\centering $10512.1\pm 2.1\pm 0.9 $~\cite{PhysRevD.98.030001} $10513.42\pm 0.41 \pm  0.18$~\cite{Sirunyan:2018dff}} \\ &&&&\\
$M_{\chi_{b2}(nP)}(MeV)$  & $9923^{+13}_{-14}$  & $10274^{+11}_{-12}$  & $10557^{+22}_{-42}$   & $\phantom{0}9912.21\pm 0.26\pm0.31$ & $\phantom{0}10268.65\pm 0.22\pm0.50$  & $10 524.02 \pm 0.57  \pm  0.18$~\cite{Sirunyan:2018dff}                    \\ \hline
\multirow{2}{*} {\centering $\bar{M}_{nP}(MeV)$  }  &\multirow{2}{*} {\centering $9906^{+8}_{-10}$ }   &\multirow{2}{*} {\centering $10263^{+8}_{-11}$ } &\multirow{2}{*} {\centering $10549^{+23}_{-41}$     }&\multirow{2}{*} {\centering $\phantom{0}9899.87 \pm 0.27$} & \multirow{2}{*} {\centering$\phantom{0}10260.20 \pm 0.36$ }& \multirow{2}{3.3cm}{\centering $10530 \pm 5\pm 9$~\cite{Aad:2011ih} $10551\pm 14 \pm 17$~\cite{Abazov:2012gh}}    \\
&&&&
\end{tabular}
\end{ruledtabular}
\end{table*}

Finally,  we would like to mention that it has been theorized that some of
the states in the  $\chi_{b}(3P)$ system could be the bottomonia
counterparts of the $X(3872)$ charmonium~\cite{Karliner:2014lta, Ferretti:2014xqa},
i.e. states closely related to the opening if the $B\bar{B}$, $BB^*$, and $B_sB_s$
thresholds.
Our results do not support such hypothesis, as the model reproduces the
$\chi_{b}(3P)$ system within (large) uncertainties,
contrary to the $X(3872)$ case which was overestimated 
using the same model~\cite{molina:2017iaa},
and whose description (both mass and width) 
calls for additional dynamics beyond the $c\bar{c}$ picture.
Along the same ideas, according to Ref.~\cite{Zhou:2018hlv}, 
the $\chi_{b1}(4P)$ state could significantly couple to the 
$B\bar{B}^*$ and $B^*\bar{B}^*$ channels.
The measurement of this particular state 
combined with the comparison to quark model calculations,
like the one presented in this work, can provide insight on the impact in the masses 
of the dynamical effects due to the open bottom thresholds.

\begin{table}
\centering
\caption{Differences $\Delta S_n=n^3S-n^1S$. 
We observe that these differences decrease when $n$ is increased.
All the differences reported in this Table are in MeV. These values have been obtained by mean of bootstrap technique. The experimental errors, in the fourth column, has been summed in quadrature.}

\label{tab:table8}
\begin{ruledtabular}
\begin{tabular}{c|c|c|c}
\multicolumn{1}{l|}{\multirow{2}{*}{$\Delta S_n$}} & \multicolumn{2}{c|}{Theory}                                 & \multirow{2}{*}{Experiment} \\ \cline{2-3}
\multicolumn{1}{l|}{}                            & \multicolumn{1}{c|}{Potential I} & \multicolumn{1}{c|}{Potential II} &                      \\ \hline
$\Delta S_1$                                     & $53^{+10}_{-18}$            & $51^{+16}_{-13}$           & $61.3\pm 2.3$               \\ 
$\Delta S_2$                                     & $18^{+9}_{-10}$              & $17^{+31}_{-5}$             & $24.3\pm4.5$              \\ 
$\Delta S_3$                                     & $11^{+4}_{-31}$             & $11^{+26}_{-4}$            & $---$                \\ 
$\Delta S_4$                                     &  $8^{+11}_{-54}$            & $8^{+13}_{-3}$            &  $---$      \\
$\Delta S_5$                                     &  $6^{+22}_{-67}$            & $6^{+19}_{-8}$            & $---$    \\ 
\end{tabular}
\end{ruledtabular}
\end{table}

\subsection{Missing resonances}\label{miss_res}
Besides reproducing the experimentally established states, 
in Table~\ref{tab:table7} we provide predictions of states
both above and below the open bottom thresholds  ($\approx 10.6$ GeV).
In total, we predict 38  states up to 11.3 GeV 
for $0, 1, 2$ (with either $\pm$ combinations for $P$ and $C$) 
and $3^{--}$ quantum numbers.
These predicted states are of interest for future analysis at LHCb~\cite{Godfrey:2015dia,Bediaga:2018lhg,Hu:2017pat}  and Belle II~\cite{Fulsom:2017erj,Ye:2016pgb,Kou:2018nap,Tanida:2019yvl,Pedlar:2018tck}.
In particular,  pinning down the $\Upsilon(6S)$ would provide further insight on 
bottomonium-like states~\cite{Kou:2018nap}.

The missing $\eta_b(nS)$ sector ($n^1S_0$ states)
can be studied through their relation to their angular momentum partners
$\Upsilon(nS)$  ($n^3S_1$) --known from experiment--, 
by computing the  $\Delta S_n=n^3S-n^1S$ mass splitting.
This difference should decrease as $n$ increases in the potential model context~\cite{Olsen:2014qna}. The 
experimental data for $\Delta S_1$ and $\Delta S_2$ shown in Table~\ref{tab:table8} support this theoretical results.
Thereby, we consider our mass estimations for both 
$\eta_b(nS)$ and $\Upsilon(nS)$ reasonable.

We also provide predictions for states of the  $n^{1,3}D_{1,2,3}$ family, which remain undetected
except for the $1^3D_2$ resonance.
The predicted missing states (with uncertainties) provide useful information to guide the
forthcoming spectroscopy programs in 
 Belle II~\cite{Tanida:2019yvl,Pedlar:2018tck} and
LHC~\cite{Bediaga:2018lhg,Piucci:2017kih}.
However, the production rate of these states should be low, hence,
difficult to detect~\cite{Godfrey:2015dia}.
 
\section{Conclusions}\label{conclu}
We have developed a relativistic quark model 
in momentum space to study the bottomonium
spectrum.
The model closely follows the one used in
Ref.~\cite{molina:2017iaa} to study charmonium.
It combines
vector and scalar interactions with a momentum dependent screening factor 
to account  for the  effects  due to  virtual  pair  creation
that  appear  close  to  the  decay  thresholds.
We fitted our model to 
all the known states of each $J^{PC}$ below the $B\bar{B}$ threshold
except for the recently measured $\chi_{b1}(3P)$ and $\chi_{b2}(3P)$
which we prefer to predict in order to gain insight on their nature
and the $\eta_b(2S)$ which we exclude of our fit
owing to the disagreement between CLEO and Belle measurements.
Our prediction for $\eta_b(2S)$ mass agrees with the Belle result.

We have performed a full statistical error analysis using the bootstrap technique, 
that provides a rigorous treatment of the statistical uncertainties.
In this way we obtain the uncertainties of the parameters and their correlations
and we can propagate both to the predicted spectrum.
Previous error analysis within phenomenological models have been very limited and incomplete. 
The rigorous error estimations allow us to assess if the inclusion of a new effect in the phenomenological model is necessary or not, and the correlations provide insight on
 how independent are the different pieces of the model among them.
 A full error analysis is mandatory to identify which deviations from experimental data can be absorbed into the statistical uncertainties of the models and which can be related to physics beyond the $b\bar{b}$ picture, guiding future research.
We find that 
the model reproduces very well the fitted states as well 
as the nonfitted ones within uncertainties.

To asses the importance of a confining term in the scalar interaction,
i.e. $\beta_s\ne 0$ in Eq.~\eqref{scalar}, we fitted 
the data with and without such contribution.
The results obtained  with the two potentials are very similar
for the fitted and the predicted states, both in the low and the high parts of the spectrum. 
Therefore,  such confining contribution to the scalar interaction can be disregarded
in a bottomonium relativistic model.
Even so, the correlations found among the parameters belonging to the scalar
interaction and the rest of the model parameters,  
show that the scalar interaction $A$ in Eq.~\eqref{scalar} 
is strictly necessary to reproduce the spectrum. 
The screening factor $F_s(p)$ included in the interaction Hamiltonian  
begins to impact the predictions in significant way at $\approx 11200$ MeV,
i.e. further away from the open bottom decay thresholds.
Hence, the screening effect is not particularly intense and
has a slight impact on the bottomonium spectrum, 
contrary to what it was found for the
charmonium one~\cite{molina:2017iaa}.

We have also studied  the $\chi_{b}(3P)$ resonances.
In particular we have calculated the mass of each  state of this system and its barycenter. 
The experimental mass value of the $\chi_{b1}(3P)$ falls 
into the theoretical uncertainty calculated with both potentials. 
Whereby, we conclude that the model is able to properly predict this state. Also, the model, with both potentials, reproduces the $\chi_{b1}(3P)$ state.  
Our result indicates that the $\chi_{b1,2}(3P)$  states
are more likely to be $b\bar{b}$ mesons than the hypothetical $X_b$ states.

Our model overestimates the $\Upsilon (4S)$ mass and is consistent with results obtained
by semirelativistic quark models, within errors. This is an indication of physics
beyond the $b\bar{b}$ picture for this state.
We identify the $\Upsilon (10860)$ as a  $5^3S_1$  state and the model fails to reproduce the
 $\Upsilon (11020)$, although it is well reproduced by other potential models
 that take into account coupled channel effects~\cite{Liu:2011yp}. Hence, the first can be considered (mostly) a $b\bar{b}$ state while the latter is up for discussion.

Finally, we report some states that, up to now, 
have not been observed experimentally but
the confirmation of their existence is part of the experimental 
plans at 
LHC $B$ factories and Belle II.

\begin{acknowledgments}
This work was supported in part by 
PAPIIT-DGAPA (UNAM, Mexico) Grant No.~IA101819
and CONACYT (Mexico) Grants No.~251817
and~No.~A1-S-21389.
\end{acknowledgments}

\bibliography{bibi2}

\begin{thebibliography}{82}%
\makeatletter
\providecommand \@ifxundefined [1]{%
 \@ifx{#1\undefined}
}%
\providecommand \@ifnum [1]{%
 \ifnum #1\expandafter \@firstoftwo
 \else \expandafter \@secondoftwo
 \fi
}%
\providecommand \@ifx [1]{%
 \ifx #1\expandafter \@firstoftwo
 \else \expandafter \@secondoftwo
 \fi
}%
\providecommand \natexlab [1]{#1}%
\providecommand \enquote  [1]{``#1''}%
\providecommand \bibnamefont  [1]{#1}%
\providecommand \bibfnamefont [1]{#1}%
\providecommand \citenamefont [1]{#1}%
\providecommand \href@noop [0]{\@secondoftwo}%
\providecommand \href [0]{\begingroup \@sanitize@url \@href}%
\providecommand \@href[1]{\@@startlink{#1}\@@href}%
\providecommand \@@href[1]{\endgroup#1\@@endlink}%
\providecommand \@sanitize@url [0]{\catcode `\\12\catcode `\$12\catcode
  `\&12\catcode `\#12\catcode `\^12\catcode `\_12\catcode `\%12\relax}%
\providecommand \@@startlink[1]{}%
\providecommand \@@endlink[0]{}%
\providecommand \url  [0]{\begingroup\@sanitize@url \@url }%
\providecommand \@url [1]{\endgroup\@href {#1}{\urlprefix }}%
\providecommand \urlprefix  [0]{URL }%
\providecommand \Eprint [0]{\href }%
\providecommand \doibase [0]{http://dx.doi.org/}%
\providecommand \selectlanguage [0]{\@gobble}%
\providecommand \bibinfo  [0]{\@secondoftwo}%
\providecommand \bibfield  [0]{\@secondoftwo}%
\providecommand \translation [1]{[#1]}%
\providecommand \BibitemOpen [0]{}%
\providecommand \bibitemStop [0]{}%
\providecommand \bibitemNoStop [0]{.\EOS\space}%
\providecommand \EOS [0]{\spacefactor3000\relax}%
\providecommand \BibitemShut  [1]{\csname bibitem#1\endcsname}%
\let\auto@bib@innerbib\@empty
\bibitem [{\citenamefont {Aad}\ \emph {et~al.}(2012)\citenamefont {Aad} \emph
  {et~al.}}]{Aad:2011ih}%
  \BibitemOpen
  \bibfield  {author} {\bibinfo {author} {\bibfnamefont {G.}~\bibnamefont
  {Aad}} \emph {et~al.} (\bibinfo {collaboration} {ATLAS}),\ }\href {\doibase
  10.1103/PhysRevLett.108.152001} {\bibfield  {journal} {\bibinfo  {journal}
  {Phys. Rev. Lett.}\ }\textbf {\bibinfo {volume} {108}},\ \bibinfo {pages}
  {152001} (\bibinfo {year} {2012})},\ \Eprint {http://arxiv.org/abs/1112.5154}
  {arXiv:1112.5154 [hep-ex]} \BibitemShut {NoStop}%
\bibitem [{\citenamefont {Abazov}\ \emph {et~al.}(2012)\citenamefont {Abazov}
  \emph {et~al.}}]{Abazov:2012gh}%
  \BibitemOpen
  \bibfield  {author} {\bibinfo {author} {\bibfnamefont {V.~M.}\ \bibnamefont
  {Abazov}} \emph {et~al.} (\bibinfo {collaboration} {D0}),\ }\href {\doibase
  10.1103/PhysRevD.86.031103} {\bibfield  {journal} {\bibinfo  {journal} {Phys.
  Rev.}\ }\textbf {\bibinfo {volume} {D86}},\ \bibinfo {pages} {031103}
  (\bibinfo {year} {2012})},\ \Eprint {http://arxiv.org/abs/1203.6034}
  {arXiv:1203.6034 [hep-ex]} \BibitemShut {NoStop}%
\bibitem [{\citenamefont {Park}(2017)}]{Park:2017pne}%
  \BibitemOpen
  \bibfield  {author} {\bibinfo {author} {\bibfnamefont {J.}~\bibnamefont
  {Park}} (\bibinfo {collaboration} {CMS}),\ }\bibfield  {booktitle} {\emph
  {\bibinfo {booktitle} {{Proceedings, 46th International Symposium on
  Multiparticle Dynamics (ISMD 2016): Jeju Island, South Korea, August
  29-September 2, 2016}}},\ }\href {\doibase 10.1051/epjconf/201714108007}
  {\bibfield  {journal} {\bibinfo  {journal} {EPJ Web Conf.}\ }\textbf
  {\bibinfo {volume} {141}},\ \bibinfo {pages} {08007} (\bibinfo {year}
  {2017})}\BibitemShut {NoStop}%
\bibitem [{\citenamefont {Sirunyan}\ \emph {et~al.}(2019)\citenamefont
  {Sirunyan} \emph {et~al.}}]{Sirunyan:2019osb}%
  \BibitemOpen
  \bibfield  {author} {\bibinfo {author} {\bibfnamefont {A.~M.}\ \bibnamefont
  {Sirunyan}} \emph {et~al.} (\bibinfo {collaboration} {CMS}),\ }\href
  {\doibase 10.1103/PhysRevLett.122.132001} {\bibfield  {journal} {\bibinfo
  {journal} {Phys. Rev. Lett.}\ }\textbf {\bibinfo {volume} {122}},\ \bibinfo
  {pages} {132001} (\bibinfo {year} {2019})},\ \Eprint
  {http://arxiv.org/abs/1902.00571} {arXiv:1902.00571 [hep-ex]} \BibitemShut
  {NoStop}%
\bibitem [{\citenamefont {Aaij}\ \emph {et~al.}(2019)\citenamefont {Aaij} \emph
  {et~al.}}]{Aaij:2019evc}%
  \BibitemOpen
  \bibfield  {author} {\bibinfo {author} {\bibfnamefont {R.}~\bibnamefont
  {Aaij}} \emph {et~al.} (\bibinfo {collaboration} {LHCb}),\ }\href {\doibase
  10.1007/JHEP07(2019)035} {\bibfield  {journal} {\bibinfo  {journal} {JHEP}\
  }\textbf {\bibinfo {volume} {07}},\ \bibinfo {pages} {035} (\bibinfo {year}
  {2019})},\ \Eprint {http://arxiv.org/abs/1903.12240} {arXiv:1903.12240
  [hep-ex]} \BibitemShut {NoStop}%
\bibitem [{\citenamefont {Fulsom}\ \emph {et~al.}(2018)\citenamefont {Fulsom}
  \emph {et~al.}}]{Fulsom:2018hpf}%
  \BibitemOpen
  \bibfield  {author} {\bibinfo {author} {\bibfnamefont {B.~G.}\ \bibnamefont
  {Fulsom}} \emph {et~al.} (\bibinfo {collaboration} {Belle}),\ }\href
  {\doibase 10.1103/PhysRevLett.121.232001} {\bibfield  {journal} {\bibinfo
  {journal} {Phys. Rev. Lett.}\ }\textbf {\bibinfo {volume} {121}},\ \bibinfo
  {pages} {232001} (\bibinfo {year} {2018})},\ \Eprint
  {http://arxiv.org/abs/1807.01201} {arXiv:1807.01201 [hep-ex]} \BibitemShut
  {NoStop}%
\bibitem [{\citenamefont {Sirunyan}\ \emph {et~al.}(2018)\citenamefont
  {Sirunyan} \emph {et~al.}}]{Sirunyan:2018dff}%
  \BibitemOpen
  \bibfield  {author} {\bibinfo {author} {\bibfnamefont {A.~M.}\ \bibnamefont
  {Sirunyan}} \emph {et~al.} (\bibinfo {collaboration} {CMS}),\ }\href
  {\doibase 10.1103/PhysRevLett.121.092002} {\bibfield  {journal} {\bibinfo
  {journal} {Phys. Rev. Lett.}\ }\textbf {\bibinfo {volume} {121}},\ \bibinfo
  {pages} {092002} (\bibinfo {year} {2018})},\ \Eprint
  {http://arxiv.org/abs/1805.11192} {arXiv:1805.11192 [hep-ex]} \BibitemShut
  {NoStop}%
\bibitem [{\citenamefont {Tamponi}\ \emph {et~al.}(2018)\citenamefont {Tamponi}
  \emph {et~al.}}]{Tamponi:2018cuf}%
  \BibitemOpen
  \bibfield  {author} {\bibinfo {author} {\bibfnamefont {U.}~\bibnamefont
  {Tamponi}} \emph {et~al.} (\bibinfo {collaboration} {Belle}),\ }\href
  {\doibase 10.1140/epjc/s10052-018-6086-4} {\bibfield  {journal} {\bibinfo
  {journal} {Eur. Phys. J.}\ }\textbf {\bibinfo {volume} {C78}},\ \bibinfo
  {pages} {633} (\bibinfo {year} {2018})},\ \Eprint
  {http://arxiv.org/abs/1803.03225} {arXiv:1803.03225 [hep-ex]} \BibitemShut
  {NoStop}%
\bibitem [{\citenamefont {Ablikim}\ \emph
  {et~al.}(2017{\natexlab{a}})\citenamefont {Ablikim} \emph
  {et~al.}}]{BESIII:2016adj}%
  \BibitemOpen
  \bibfield  {author} {\bibinfo {author} {\bibfnamefont {M.}~\bibnamefont
  {Ablikim}} \emph {et~al.} (\bibinfo {collaboration} {BESIII}),\ }\href
  {\doibase 10.1103/PhysRevLett.118.092002} {\bibfield  {journal} {\bibinfo
  {journal} {Phys. Rev. Lett.}\ }\textbf {\bibinfo {volume} {118}},\ \bibinfo
  {pages} {092002} (\bibinfo {year} {2017}{\natexlab{a}})},\ \Eprint
  {http://arxiv.org/abs/1610.07044} {arXiv:1610.07044 [hep-ex]} \BibitemShut
  {NoStop}%
\bibitem [{\citenamefont {Ablikim}\ \emph
  {et~al.}(2017{\natexlab{b}})\citenamefont {Ablikim} \emph
  {et~al.}}]{Ablikim:2016qzw}%
  \BibitemOpen
  \bibfield  {author} {\bibinfo {author} {\bibfnamefont {M.}~\bibnamefont
  {Ablikim}} \emph {et~al.} (\bibinfo {collaboration} {BESIII}),\ }\href
  {\doibase 10.1103/PhysRevLett.118.092001} {\bibfield  {journal} {\bibinfo
  {journal} {Phys. Rev. Lett.}\ }\textbf {\bibinfo {volume} {118}},\ \bibinfo
  {pages} {092001} (\bibinfo {year} {2017}{\natexlab{b}})},\ \Eprint
  {http://arxiv.org/abs/1611.01317} {arXiv:1611.01317 [hep-ex]} \BibitemShut
  {NoStop}%
\bibitem [{\citenamefont {Olsen}\ \emph {et~al.}(2018)\citenamefont {Olsen},
  \citenamefont {Skwarnicki},\ and\ \citenamefont {Zieminska}}]{Olsen:2017bmm}%
  \BibitemOpen
  \bibfield  {author} {\bibinfo {author} {\bibfnamefont {S.~L.}\ \bibnamefont
  {Olsen}}, \bibinfo {author} {\bibfnamefont {T.}~\bibnamefont {Skwarnicki}}, \
  and\ \bibinfo {author} {\bibfnamefont {D.}~\bibnamefont {Zieminska}},\ }\href
  {\doibase 10.1103/RevModPhys.90.015003} {\bibfield  {journal} {\bibinfo
  {journal} {Rev. Mod. Phys.}\ }\textbf {\bibinfo {volume} {90}},\ \bibinfo
  {pages} {015003} (\bibinfo {year} {2018})},\ \Eprint
  {http://arxiv.org/abs/1708.04012} {arXiv:1708.04012 [hep-ph]} \BibitemShut
  {NoStop}%
\bibitem [{\citenamefont {Tanabashi}(2018)}]{PhysRevD.98.030001}%
  \BibitemOpen
  \bibfield  {author} {\bibinfo {author} {\bibfnamefont {M.}~\bibnamefont
  {Tanabashi}} (\bibinfo {collaboration} {Particle Data Group}),\ }\href
  {\doibase 10.1103/PhysRevD.98.030001} {\bibfield  {journal} {\bibinfo
  {journal} {Phys. Rev.}\ }\textbf {\bibinfo {volume} {D98}},\ \bibinfo {pages}
  {030001} (\bibinfo {year} {2018})}\BibitemShut {NoStop}%
\bibitem [{\citenamefont {Tanida}(2019)}]{Tanida:2019yvl}%
  \BibitemOpen
  \bibfield  {author} {\bibinfo {author} {\bibfnamefont {K.}~\bibnamefont
  {Tanida}} (\bibinfo {collaboration} {Belle II}),\ }\bibfield  {booktitle}
  {\emph {\bibinfo {booktitle} {{Proceedings, 13th International Conference on
  Hypernuclear and Strange Particle Physics (HYP 2018): Portsmouth Virginia,
  USA, June 24-29, 2018}}},\ }\href {\doibase 10.1063/1.5118417} {\bibfield
  {journal} {\bibinfo  {journal} {AIP Conf. Proc.}\ }\textbf {\bibinfo {volume}
  {2130}},\ \bibinfo {pages} {040020} (\bibinfo {year} {2019})}\BibitemShut
  {NoStop}%
\bibitem [{\citenamefont {Pedlar}(2018)}]{Pedlar:2018tck}%
  \BibitemOpen
  \bibfield  {author} {\bibinfo {author} {\bibfnamefont {T.~K.}\ \bibnamefont
  {Pedlar}},\ }\bibfield  {booktitle} {\emph {\bibinfo {booktitle}
  {{Proceedings, 6th International Conference on Exotic Atoms and Related
  Topics (EXA2017): Vienna, Austria, September 11-15, 2017}}},\ }\href
  {\doibase 10.1051/epjconf/201818101022} {\bibfield  {journal} {\bibinfo
  {journal} {EPJ Web Conf.}\ }\textbf {\bibinfo {volume} {181}},\ \bibinfo
  {pages} {01022} (\bibinfo {year} {2018})}\BibitemShut {NoStop}%
\bibitem [{\citenamefont {Aaij}\ \emph {et~al.}(2018)\citenamefont {Aaij} \emph
  {et~al.}}]{Bediaga:2018lhg}%
  \BibitemOpen
  \bibfield  {author} {\bibinfo {author} {\bibfnamefont {R.}~\bibnamefont
  {Aaij}} \emph {et~al.} (\bibinfo {collaboration} {LHCb}),\ }\href@noop {}
  {\enquote {\bibinfo {title} {{Physics case for an LHCb Upgrade II -
  Opportunities in flavour physics, and beyond, in the HL-LHC era}},}\ }
  (\bibinfo {year} {2018}),\ \Eprint {http://arxiv.org/abs/1808.08865}
  {arXiv:1808.08865} \BibitemShut {NoStop}%
\bibitem [{\citenamefont {Piucci}(2017)}]{Piucci:2017kih}%
  \BibitemOpen
  \bibfield  {author} {\bibinfo {author} {\bibfnamefont {A.}~\bibnamefont
  {Piucci}},\ }\bibfield  {booktitle} {\emph {\bibinfo {booktitle}
  {{Proceedings, Workshop on Discovery Physics at the LHC (Kruger2016): Kruger
  National Park, Mpumalanga, South Africa, December 5-9, 2016}}},\ }\href
  {\doibase 10.1088/1742-6596/878/1/012012} {\bibfield  {journal} {\bibinfo
  {journal} {J. Phys. Conf. Ser.}\ }\textbf {\bibinfo {volume} {878}},\
  \bibinfo {pages} {012012} (\bibinfo {year} {2017})}\BibitemShut {NoStop}%
\bibitem [{\citenamefont {Guo}\ \emph {et~al.}(2018)\citenamefont {Guo},
  \citenamefont {Hanhart}, \citenamefont {Meißner}, \citenamefont {Wang},
  \citenamefont {Zhao},\ and\ \citenamefont {Zou}}]{Guo:2017jvc}%
  \BibitemOpen
  \bibfield  {author} {\bibinfo {author} {\bibfnamefont {F.-K.}\ \bibnamefont
  {Guo}}, \bibinfo {author} {\bibfnamefont {C.}~\bibnamefont {Hanhart}},
  \bibinfo {author} {\bibfnamefont {U.-G.}\ \bibnamefont {Meißner}}, \bibinfo
  {author} {\bibfnamefont {Q.}~\bibnamefont {Wang}}, \bibinfo {author}
  {\bibfnamefont {Q.}~\bibnamefont {Zhao}}, \ and\ \bibinfo {author}
  {\bibfnamefont {B.-S.}\ \bibnamefont {Zou}},\ }\href {\doibase
  10.1103/RevModPhys.90.015004} {\bibfield  {journal} {\bibinfo  {journal}
  {Rev. Mod. Phys.}\ }\textbf {\bibinfo {volume} {90}},\ \bibinfo {pages}
  {015004} (\bibinfo {year} {2018})},\ \Eprint
  {http://arxiv.org/abs/1705.00141} {arXiv:1705.00141 [hep-ph]} \BibitemShut
  {NoStop}%
\bibitem [{\citenamefont {Lebed}\ \emph {et~al.}(2017)\citenamefont {Lebed},
  \citenamefont {Mitchell},\ and\ \citenamefont {Swanson}}]{Lebed:2016hpi}%
  \BibitemOpen
  \bibfield  {author} {\bibinfo {author} {\bibfnamefont {R.~F.}\ \bibnamefont
  {Lebed}}, \bibinfo {author} {\bibfnamefont {R.~E.}\ \bibnamefont {Mitchell}},
  \ and\ \bibinfo {author} {\bibfnamefont {E.~S.}\ \bibnamefont {Swanson}},\
  }\href {\doibase 10.1016/j.ppnp.2016.11.003} {\bibfield  {journal} {\bibinfo
  {journal} {Prog. Part. Nucl. Phys.}\ }\textbf {\bibinfo {volume} {93}},\
  \bibinfo {pages} {143} (\bibinfo {year} {2017})},\ \Eprint
  {http://arxiv.org/abs/1610.04528} {arXiv:1610.04528 [hep-ph]} \BibitemShut
  {NoStop}%
\bibitem [{\citenamefont {Yuan}(2018)}]{Yuan:2018inv}%
  \BibitemOpen
  \bibfield  {author} {\bibinfo {author} {\bibfnamefont {C.-Z.}\ \bibnamefont
  {Yuan}},\ }\href {\doibase 10.1142/S0217751X18300181} {\bibfield  {journal}
  {\bibinfo  {journal} {Int. J. Mod. Phys.}\ }\textbf {\bibinfo {volume}
  {A33}},\ \bibinfo {pages} {1830018} (\bibinfo {year} {2018})},\ \Eprint
  {http://arxiv.org/abs/1808.01570} {arXiv:1808.01570 [hep-ex]} \BibitemShut
  {NoStop}%
\bibitem [{\citenamefont {Liu}\ \emph {et~al.}(2019)\citenamefont {Liu},
  \citenamefont {Chen}, \citenamefont {Chen}, \citenamefont {Liu},\ and\
  \citenamefont {Zhu}}]{Liu:2019zoy}%
  \BibitemOpen
  \bibfield  {author} {\bibinfo {author} {\bibfnamefont {Y.-R.}\ \bibnamefont
  {Liu}}, \bibinfo {author} {\bibfnamefont {H.-X.}\ \bibnamefont {Chen}},
  \bibinfo {author} {\bibfnamefont {W.}~\bibnamefont {Chen}}, \bibinfo {author}
  {\bibfnamefont {X.}~\bibnamefont {Liu}}, \ and\ \bibinfo {author}
  {\bibfnamefont {S.-L.}\ \bibnamefont {Zhu}},\ }\href {\doibase
  10.1016/j.ppnp.2019.04.003} {\bibfield  {journal} {\bibinfo  {journal} {Prog.
  Part. Nucl. Phys.}\ }\textbf {\bibinfo {volume} {107}},\ \bibinfo {pages}
  {237} (\bibinfo {year} {2019})},\ \Eprint {http://arxiv.org/abs/1903.11976}
  {arXiv:1903.11976 [hep-ph]} \BibitemShut {NoStop}%
\bibitem [{\citenamefont {Brambilla}\ \emph {et~al.}(2019)\citenamefont
  {Brambilla}, \citenamefont {Eidelman}, \citenamefont {Hanhart}, \citenamefont
  {Nefediev}, \citenamefont {Shen}, \citenamefont {Thomas}, \citenamefont
  {Vairo},\ and\ \citenamefont {Yuan}}]{Brambilla:2019esw}%
  \BibitemOpen
  \bibfield  {author} {\bibinfo {author} {\bibfnamefont {N.}~\bibnamefont
  {Brambilla}}, \bibinfo {author} {\bibfnamefont {S.}~\bibnamefont {Eidelman}},
  \bibinfo {author} {\bibfnamefont {C.}~\bibnamefont {Hanhart}}, \bibinfo
  {author} {\bibfnamefont {A.}~\bibnamefont {Nefediev}}, \bibinfo {author}
  {\bibfnamefont {C.-P.}\ \bibnamefont {Shen}}, \bibinfo {author}
  {\bibfnamefont {C.~E.}\ \bibnamefont {Thomas}}, \bibinfo {author}
  {\bibfnamefont {A.}~\bibnamefont {Vairo}}, \ and\ \bibinfo {author}
  {\bibfnamefont {C.-Z.}\ \bibnamefont {Yuan}},\ }\href@noop {} {\enquote
  {\bibinfo {title} {{The $XYZ$ states: experimental and theoretical status and
  perspectives}},}\ } (\bibinfo {year} {2019}),\ \Eprint
  {http://arxiv.org/abs/1907.07583} {arXiv:1907.07583 [hep-ex]} \BibitemShut
  {NoStop}%
\bibitem [{\citenamefont {Liao}\ and\ \citenamefont
  {Manke}(2002)}]{Liao:2001yh}%
  \BibitemOpen
  \bibfield  {author} {\bibinfo {author} {\bibfnamefont {X.}~\bibnamefont
  {Liao}}\ and\ \bibinfo {author} {\bibfnamefont {T.}~\bibnamefont {Manke}},\
  }\href {\doibase 10.1103/PhysRevD.65.074508} {\bibfield  {journal} {\bibinfo
  {journal} {Phys. Rev.}\ }\textbf {\bibinfo {volume} {D65}},\ \bibinfo {pages}
  {074508} (\bibinfo {year} {2002})},\ \Eprint
  {http://arxiv.org/abs/hep-lat/0111049} {arXiv:hep-lat/0111049 [hep-lat]}
  \BibitemShut {NoStop}%
\bibitem [{\citenamefont {Meinel}(2009)}]{Meinel:2009rd}%
  \BibitemOpen
  \bibfield  {author} {\bibinfo {author} {\bibfnamefont {S.}~\bibnamefont
  {Meinel}},\ }\href {\doibase 10.1103/PhysRevD.79.094501} {\bibfield
  {journal} {\bibinfo  {journal} {Phys. Rev.}\ }\textbf {\bibinfo {volume}
  {D79}},\ \bibinfo {pages} {094501} (\bibinfo {year} {2009})},\ \Eprint
  {http://arxiv.org/abs/0903.3224} {arXiv:0903.3224 [hep-lat]} \BibitemShut
  {NoStop}%
\bibitem [{\citenamefont {Meinel}(2010)}]{Meinel:2010pv}%
  \BibitemOpen
  \bibfield  {author} {\bibinfo {author} {\bibfnamefont {S.}~\bibnamefont
  {Meinel}},\ }\href {\doibase 10.1103/PhysRevD.82.114502} {\bibfield
  {journal} {\bibinfo  {journal} {Phys. Rev.}\ }\textbf {\bibinfo {volume}
  {D82}},\ \bibinfo {pages} {114502} (\bibinfo {year} {2010})},\ \Eprint
  {http://arxiv.org/abs/1007.3966} {arXiv:1007.3966 [hep-lat]} \BibitemShut
  {NoStop}%
\bibitem [{\citenamefont {Daldrop}\ \emph {et~al.}(2012)\citenamefont
  {Daldrop}, \citenamefont {Davies},\ and\ \citenamefont
  {Dowdall}}]{Daldrop:2011aa}%
  \BibitemOpen
  \bibfield  {author} {\bibinfo {author} {\bibfnamefont {J.~O.}\ \bibnamefont
  {Daldrop}}, \bibinfo {author} {\bibfnamefont {C.~T.~H.}\ \bibnamefont
  {Davies}}, \ and\ \bibinfo {author} {\bibfnamefont {R.~J.}\ \bibnamefont
  {Dowdall}} (\bibinfo {collaboration} {HPQCD}),\ }\href {\doibase
  10.1103/PhysRevLett.108.102003} {\bibfield  {journal} {\bibinfo  {journal}
  {Phys. Rev. Lett.}\ }\textbf {\bibinfo {volume} {108}},\ \bibinfo {pages}
  {102003} (\bibinfo {year} {2012})},\ \Eprint {http://arxiv.org/abs/1112.2590}
  {arXiv:1112.2590 [hep-lat]} \BibitemShut {NoStop}%
\bibitem [{\citenamefont {Lewis}\ and\ \citenamefont
  {Woloshyn}(2012)}]{Lewis:2012ir}%
  \BibitemOpen
  \bibfield  {author} {\bibinfo {author} {\bibfnamefont {R.}~\bibnamefont
  {Lewis}}\ and\ \bibinfo {author} {\bibfnamefont {R.~M.}\ \bibnamefont
  {Woloshyn}},\ }\href {\doibase 10.1103/PhysRevD.85.114509} {\bibfield
  {journal} {\bibinfo  {journal} {Phys. Rev.}\ }\textbf {\bibinfo {volume}
  {D85}},\ \bibinfo {pages} {114509} (\bibinfo {year} {2012})},\ \Eprint
  {http://arxiv.org/abs/1204.4675} {arXiv:1204.4675 [hep-lat]} \BibitemShut
  {NoStop}%
\bibitem [{\citenamefont {Dowdall}\ \emph {et~al.}(2014)\citenamefont
  {Dowdall}, \citenamefont {Davies}, \citenamefont {Hammant}, \citenamefont
  {Horgan},\ and\ \citenamefont {Hughes}}]{Dowdall:2013jqa}%
  \BibitemOpen
  \bibfield  {author} {\bibinfo {author} {\bibfnamefont {R.~J.}\ \bibnamefont
  {Dowdall}}, \bibinfo {author} {\bibfnamefont {C.~T.~H.}\ \bibnamefont
  {Davies}}, \bibinfo {author} {\bibfnamefont {T.}~\bibnamefont {Hammant}},
  \bibinfo {author} {\bibfnamefont {R.~R.}\ \bibnamefont {Horgan}}, \ and\
  \bibinfo {author} {\bibfnamefont {C.}~\bibnamefont {Hughes}} (\bibinfo
  {collaboration} {HPQCD}),\ }\href {\doibase 10.1103/PhysRevD.92.039904,
  10.1103/PhysRevD.89.031502} {\bibfield  {journal} {\bibinfo  {journal} {Phys.
  Rev.}\ }\textbf {\bibinfo {volume} {D89}},\ \bibinfo {pages} {031502}
  (\bibinfo {year} {2014})},\ \bibinfo {note} {[Erratum: Phys.
  Rev.D92,039904(2015)]},\ \Eprint {http://arxiv.org/abs/1309.5797}
  {arXiv:1309.5797 [hep-lat]} \BibitemShut {NoStop}%
\bibitem [{\citenamefont {Davies}\ \emph {et~al.}(2014)\citenamefont {Davies},
  \citenamefont {Colquhoun}, \citenamefont {Galloway}, \citenamefont {Donald},
  \citenamefont {Koponen}, \citenamefont {Dowdall}, \citenamefont {Horgan},
  \citenamefont {Follana}, \citenamefont {Lepage},\ and\ \citenamefont
  {McNeile}}]{Davies:2013dem}%
  \BibitemOpen
  \bibfield  {author} {\bibinfo {author} {\bibfnamefont {C.~T.~H.}\
  \bibnamefont {Davies}}, \bibinfo {author} {\bibfnamefont {B.}~\bibnamefont
  {Colquhoun}}, \bibinfo {author} {\bibfnamefont {B.}~\bibnamefont {Galloway}},
  \bibinfo {author} {\bibfnamefont {G.~C.}\ \bibnamefont {Donald}}, \bibinfo
  {author} {\bibfnamefont {J.}~\bibnamefont {Koponen}}, \bibinfo {author}
  {\bibfnamefont {R.~J.}\ \bibnamefont {Dowdall}}, \bibinfo {author}
  {\bibfnamefont {R.}~\bibnamefont {Horgan}}, \bibinfo {author} {\bibfnamefont
  {E.}~\bibnamefont {Follana}}, \bibinfo {author} {\bibfnamefont {G.~P.}\
  \bibnamefont {Lepage}}, \ and\ \bibinfo {author} {\bibfnamefont
  {C.}~\bibnamefont {McNeile}} (\bibinfo {collaboration} {HPQCD}),\ }\bibfield
  {booktitle} {\emph {\bibinfo {booktitle} {{Proceedings, 31st International
  Symposium on Lattice Field Theory (Lattice 2013): Mainz, Germany, July
  29-August 3, 2013}}},\ }\href {\doibase 10.22323/1.187.0438} {\bibfield
  {journal} {\bibinfo  {journal} {PoS}\ }\textbf {\bibinfo {volume}
  {LATTICE2013}},\ \bibinfo {pages} {438} (\bibinfo {year} {2014})},\ \Eprint
  {http://arxiv.org/abs/1312.5874} {arXiv:1312.5874 [hep-lat]} \BibitemShut
  {NoStop}%
\bibitem [{\citenamefont {Baker}\ \emph {et~al.}(2015)\citenamefont {Baker},
  \citenamefont {Penin}, \citenamefont {Seidel},\ and\ \citenamefont
  {Zerf}}]{Baker:2015xma}%
  \BibitemOpen
  \bibfield  {author} {\bibinfo {author} {\bibfnamefont {M.}~\bibnamefont
  {Baker}}, \bibinfo {author} {\bibfnamefont {A.~A.}\ \bibnamefont {Penin}},
  \bibinfo {author} {\bibfnamefont {D.}~\bibnamefont {Seidel}}, \ and\ \bibinfo
  {author} {\bibfnamefont {N.}~\bibnamefont {Zerf}},\ }\href {\doibase
  10.1103/PhysRevD.92.054502} {\bibfield  {journal} {\bibinfo  {journal} {Phys.
  Rev.}\ }\textbf {\bibinfo {volume} {D92}},\ \bibinfo {pages} {054502}
  (\bibinfo {year} {2015})},\ \Eprint {http://arxiv.org/abs/1504.05979}
  {arXiv:1504.05979 [hep-ph]} \BibitemShut {NoStop}%
\bibitem [{\citenamefont {Ding}\ \emph {et~al.}(2019)\citenamefont {Ding},
  \citenamefont {Kaczmarek}, \citenamefont {Kruse}, \citenamefont {Larsen},
  \citenamefont {Mazur}, \citenamefont {Mukherjee}, \citenamefont {Ohno},
  \citenamefont {Sandmeyer},\ and\ \citenamefont {Shu}}]{Ding:2018uhl}%
  \BibitemOpen
  \bibfield  {author} {\bibinfo {author} {\bibfnamefont {H.-T.}\ \bibnamefont
  {Ding}}, \bibinfo {author} {\bibfnamefont {O.}~\bibnamefont {Kaczmarek}},
  \bibinfo {author} {\bibfnamefont {A.-L.}\ \bibnamefont {Kruse}}, \bibinfo
  {author} {\bibfnamefont {R.}~\bibnamefont {Larsen}}, \bibinfo {author}
  {\bibfnamefont {L.}~\bibnamefont {Mazur}}, \bibinfo {author} {\bibfnamefont
  {S.}~\bibnamefont {Mukherjee}}, \bibinfo {author} {\bibfnamefont
  {H.}~\bibnamefont {Ohno}}, \bibinfo {author} {\bibfnamefont {H.}~\bibnamefont
  {Sandmeyer}}, \ and\ \bibinfo {author} {\bibfnamefont {H.-T.}\ \bibnamefont
  {Shu}},\ }\bibfield  {booktitle} {\emph {\bibinfo {booktitle} {{Proceedings,
  27th International Conference on Ultrarelativistic Nucleus-Nucleus Collisions
  (Quark Matter 2018): Venice, Italy, May 14-19, 2018}}},\ }\href {\doibase
  10.1016/j.nuclphysa.2018.09.075} {\bibfield  {journal} {\bibinfo  {journal}
  {Nucl. Phys.}\ }\textbf {\bibinfo {volume} {A982}},\ \bibinfo {pages} {715}
  (\bibinfo {year} {2019})},\ \Eprint {http://arxiv.org/abs/1807.06315}
  {arXiv:1807.06315 [hep-lat]} \BibitemShut {NoStop}%
\bibitem [{\citenamefont {Larsen}\ \emph {et~al.}(2020)\citenamefont {Larsen},
  \citenamefont {Meinel}, \citenamefont {Mukherjee},\ and\ \citenamefont
  {Petreczky}}]{Larsen:2019zqv}%
  \BibitemOpen
  \bibfield  {author} {\bibinfo {author} {\bibfnamefont {R.}~\bibnamefont
  {Larsen}}, \bibinfo {author} {\bibfnamefont {S.}~\bibnamefont {Meinel}},
  \bibinfo {author} {\bibfnamefont {S.}~\bibnamefont {Mukherjee}}, \ and\
  \bibinfo {author} {\bibfnamefont {P.}~\bibnamefont {Petreczky}},\ }\href
  {\doibase 10.1016/j.physletb.2019.135119} {\bibfield  {journal} {\bibinfo
  {journal} {Phys. Lett.}\ }\textbf {\bibinfo {volume} {B800}},\ \bibinfo
  {pages} {135119} (\bibinfo {year} {2020})},\ \Eprint
  {http://arxiv.org/abs/1910.07374} {arXiv:1910.07374 [hep-lat]} \BibitemShut
  {NoStop}%
\bibitem [{\citenamefont {Bicudo}\ \emph {et~al.}(2019)\citenamefont {Bicudo},
  \citenamefont {Cardoso}, \citenamefont {Cardoso},\ and\ \citenamefont
  {Wagner}}]{Bicudo:2019ymo}%
  \BibitemOpen
  \bibfield  {author} {\bibinfo {author} {\bibfnamefont {P.}~\bibnamefont
  {Bicudo}}, \bibinfo {author} {\bibfnamefont {M.}~\bibnamefont {Cardoso}},
  \bibinfo {author} {\bibfnamefont {N.}~\bibnamefont {Cardoso}}, \ and\
  \bibinfo {author} {\bibfnamefont {M.}~\bibnamefont {Wagner}},\ }\href@noop {}
  {\  (\bibinfo {year} {2019})},\ \Eprint {http://arxiv.org/abs/1910.04827}
  {arXiv:1910.04827 [hep-lat]} \BibitemShut {NoStop}%
\bibitem [{\citenamefont {Larsen}\ \emph {et~al.}(2019)\citenamefont {Larsen},
  \citenamefont {Meinel}, \citenamefont {Mukherjee},\ and\ \citenamefont
  {Petreczky}}]{Larsen:2019bwy}%
  \BibitemOpen
  \bibfield  {author} {\bibinfo {author} {\bibfnamefont {R.}~\bibnamefont
  {Larsen}}, \bibinfo {author} {\bibfnamefont {S.}~\bibnamefont {Meinel}},
  \bibinfo {author} {\bibfnamefont {S.}~\bibnamefont {Mukherjee}}, \ and\
  \bibinfo {author} {\bibfnamefont {P.}~\bibnamefont {Petreczky}},\ }\href
  {\doibase 10.1103/PhysRevD.100.074506} {\bibfield  {journal} {\bibinfo
  {journal} {Phys. Rev.}\ }\textbf {\bibinfo {volume} {D100}},\ \bibinfo
  {pages} {074506} (\bibinfo {year} {2019})},\ \Eprint
  {http://arxiv.org/abs/1908.08437} {arXiv:1908.08437 [hep-lat]} \BibitemShut
  {NoStop}%
\bibitem [{\citenamefont {Fischer}\ \emph {et~al.}(2015)\citenamefont
  {Fischer}, \citenamefont {Kubrak},\ and\ \citenamefont
  {Williams}}]{Fischer:2014cfa}%
  \BibitemOpen
  \bibfield  {author} {\bibinfo {author} {\bibfnamefont {C.~S.}\ \bibnamefont
  {Fischer}}, \bibinfo {author} {\bibfnamefont {S.}~\bibnamefont {Kubrak}}, \
  and\ \bibinfo {author} {\bibfnamefont {R.}~\bibnamefont {Williams}},\ }\href
  {\doibase 10.1140/epja/i2015-15010-7} {\bibfield  {journal} {\bibinfo
  {journal} {Eur. Phys. J.}\ }\textbf {\bibinfo {volume} {A51}},\ \bibinfo
  {pages} {10} (\bibinfo {year} {2015})},\ \Eprint
  {http://arxiv.org/abs/1409.5076} {arXiv:1409.5076 [hep-ph]} \BibitemShut
  {NoStop}%
\bibitem [{\citenamefont {Popovici}\ \emph {et~al.}(2015)\citenamefont
  {Popovici}, \citenamefont {Hilger}, \citenamefont {Gómez-Rocha},\ and\
  \citenamefont {Krassnigg}}]{Popovici:2014pha}%
  \BibitemOpen
  \bibfield  {author} {\bibinfo {author} {\bibfnamefont {C.}~\bibnamefont
  {Popovici}}, \bibinfo {author} {\bibfnamefont {T.}~\bibnamefont {Hilger}},
  \bibinfo {author} {\bibfnamefont {M.}~\bibnamefont {Gómez-Rocha}}, \ and\
  \bibinfo {author} {\bibfnamefont {A.}~\bibnamefont {Krassnigg}},\ }\bibfield
  {booktitle} {\emph {\bibinfo {booktitle} {{Proceedings, Theory and Experiment
  for Hadrons on the Light-Front (Light Cone 2014): Raleigh, North Carolina,
  USA, May 26-30, 2014}}},\ }\href {\doibase 10.1007/s00601-014-0934-z}
  {\bibfield  {journal} {\bibinfo  {journal} {Few Body Syst.}\ }\textbf
  {\bibinfo {volume} {56}},\ \bibinfo {pages} {481} (\bibinfo {year} {2015})},\
  \Eprint {http://arxiv.org/abs/1407.7970} {arXiv:1407.7970 [hep-ph]}
  \BibitemShut {NoStop}%
\bibitem [{\citenamefont {Hilger}\ \emph {et~al.}(2015)\citenamefont {Hilger},
  \citenamefont {Gomez-Rocha},\ and\ \citenamefont
  {Krassnigg}}]{Hilger:2015hka}%
  \BibitemOpen
  \bibfield  {author} {\bibinfo {author} {\bibfnamefont {T.}~\bibnamefont
  {Hilger}}, \bibinfo {author} {\bibfnamefont {M.}~\bibnamefont {Gomez-Rocha}},
  \ and\ \bibinfo {author} {\bibfnamefont {A.}~\bibnamefont {Krassnigg}},\
  }\href {\doibase 10.1103/PhysRevD.91.114004} {\bibfield  {journal} {\bibinfo
  {journal} {Phys. Rev.}\ }\textbf {\bibinfo {volume} {D91}},\ \bibinfo {pages}
  {114004} (\bibinfo {year} {2015})},\ \Eprint
  {http://arxiv.org/abs/1503.08697} {arXiv:1503.08697 [hep-ph]} \BibitemShut
  {NoStop}%
\bibitem [{\citenamefont {Negash}\ and\ \citenamefont
  {Bhatnagar}(2017)}]{Negash:2017rqt}%
  \BibitemOpen
  \bibfield  {author} {\bibinfo {author} {\bibfnamefont {H.}~\bibnamefont
  {Negash}}\ and\ \bibinfo {author} {\bibfnamefont {S.}~\bibnamefont
  {Bhatnagar}},\ }\href {\doibase 10.1155/2017/7306825} {\bibfield  {journal}
  {\bibinfo  {journal} {Adv. High Energy Phys.}\ }\textbf {\bibinfo {volume}
  {2017}},\ \bibinfo {pages} {7306825} (\bibinfo {year} {2017})},\ \Eprint
  {http://arxiv.org/abs/1703.06082} {arXiv:1703.06082 [hep-ph]} \BibitemShut
  {NoStop}%
\bibitem [{\citenamefont {Leitão}\ \emph {et~al.}(2017)\citenamefont
  {Leitão}, \citenamefont {Li}, \citenamefont {Maris}, \citenamefont {Peña},
  \citenamefont {Stadler}, \citenamefont {Vary},\ and\ \citenamefont
  {Biernat}}]{Leitao:2017esb}%
  \BibitemOpen
  \bibfield  {author} {\bibinfo {author} {\bibfnamefont {S.}~\bibnamefont
  {Leitão}}, \bibinfo {author} {\bibfnamefont {Y.}~\bibnamefont {Li}},
  \bibinfo {author} {\bibfnamefont {P.}~\bibnamefont {Maris}}, \bibinfo
  {author} {\bibfnamefont {M.~T.}\ \bibnamefont {Peña}}, \bibinfo {author}
  {\bibfnamefont {A.}~\bibnamefont {Stadler}}, \bibinfo {author} {\bibfnamefont
  {J.~P.}\ \bibnamefont {Vary}}, \ and\ \bibinfo {author} {\bibfnamefont
  {E.~P.}\ \bibnamefont {Biernat}},\ }\href {\doibase
  10.1140/epjc/s10052-017-5248-0} {\bibfield  {journal} {\bibinfo  {journal}
  {Eur. Phys. J.}\ }\textbf {\bibinfo {volume} {C77}},\ \bibinfo {pages} {696}
  (\bibinfo {year} {2017})},\ \Eprint {http://arxiv.org/abs/1705.06178}
  {arXiv:1705.06178 [hep-ph]} \BibitemShut {NoStop}%
\bibitem [{\citenamefont {Mojica}\ \emph {et~al.}(2017)\citenamefont {Mojica},
  \citenamefont {Vera}, \citenamefont {Rojas},\ and\ \citenamefont
  {El-Bennich}}]{Mojica:2017tvh}%
  \BibitemOpen
  \bibfield  {author} {\bibinfo {author} {\bibfnamefont {F.~F.}\ \bibnamefont
  {Mojica}}, \bibinfo {author} {\bibfnamefont {C.~E.}\ \bibnamefont {Vera}},
  \bibinfo {author} {\bibfnamefont {E.}~\bibnamefont {Rojas}}, \ and\ \bibinfo
  {author} {\bibfnamefont {B.}~\bibnamefont {El-Bennich}},\ }\href {\doibase
  10.1103/PhysRevD.96.014012} {\bibfield  {journal} {\bibinfo  {journal} {Phys.
  Rev.}\ }\textbf {\bibinfo {volume} {D96}},\ \bibinfo {pages} {014012}
  (\bibinfo {year} {2017})},\ \Eprint {http://arxiv.org/abs/1704.08593}
  {arXiv:1704.08593 [hep-ph]} \BibitemShut {NoStop}%
\bibitem [{\citenamefont {Wang}\ and\ \citenamefont {Wu}(2019)}]{Wang:2019tqf}%
  \BibitemOpen
  \bibfield  {author} {\bibinfo {author} {\bibfnamefont {G.-L.}\ \bibnamefont
  {Wang}}\ and\ \bibinfo {author} {\bibfnamefont {X.-G.}\ \bibnamefont {Wu}},\
  }\href@noop {} {\  (\bibinfo {year} {2019})},\ \Eprint
  {http://arxiv.org/abs/1902.06177} {arXiv:1902.06177 [hep-ph]} \BibitemShut
  {NoStop}%
\bibitem [{\citenamefont {Godfrey}\ and\ \citenamefont
  {Isgur}(1985)}]{Godfrey:1985xj}%
  \BibitemOpen
  \bibfield  {author} {\bibinfo {author} {\bibfnamefont {S.}~\bibnamefont
  {Godfrey}}\ and\ \bibinfo {author} {\bibfnamefont {N.}~\bibnamefont
  {Isgur}},\ }\href {\doibase 10.1103/PhysRevD.32.189} {\bibfield  {journal}
  {\bibinfo  {journal} {Phys. Rev.}\ }\textbf {\bibinfo {volume} {D32}},\
  \bibinfo {pages} {189} (\bibinfo {year} {1985})}\BibitemShut {NoStop}%
\bibitem [{\citenamefont {Gonzalez}(2009)}]{Gonzalez:2009jk}%
  \BibitemOpen
  \bibfield  {author} {\bibinfo {author} {\bibfnamefont {P.}~\bibnamefont
  {Gonzalez}},\ }\href {\doibase 10.1103/PhysRevD.80.054010} {\bibfield
  {journal} {\bibinfo  {journal} {Phys. Rev.}\ }\textbf {\bibinfo {volume}
  {D80}},\ \bibinfo {pages} {054010} (\bibinfo {year} {2009})},\ \Eprint
  {http://arxiv.org/abs/0909.1204} {arXiv:0909.1204 [hep-ph]} \BibitemShut
  {NoStop}%
\bibitem [{\citenamefont {Liu}\ and\ \citenamefont {Ding}(2012)}]{Liu:2011yp}%
  \BibitemOpen
  \bibfield  {author} {\bibinfo {author} {\bibfnamefont {J.-F.}\ \bibnamefont
  {Liu}}\ and\ \bibinfo {author} {\bibfnamefont {G.-J.}\ \bibnamefont {Ding}},\
  }\href {\doibase 10.1140/epjc/s10052-012-1981-6} {\bibfield  {journal}
  {\bibinfo  {journal} {Eur. Phys. J.}\ }\textbf {\bibinfo {volume} {C72}},\
  \bibinfo {pages} {1981} (\bibinfo {year} {2012})},\ \Eprint
  {http://arxiv.org/abs/1105.0855} {arXiv:1105.0855 [hep-ph]} \BibitemShut
  {NoStop}%
\bibitem [{\citenamefont {Ferretti}\ and\ \citenamefont
  {Santopinto}(2014)}]{Ferretti:2013vua}%
  \BibitemOpen
  \bibfield  {author} {\bibinfo {author} {\bibfnamefont {J.}~\bibnamefont
  {Ferretti}}\ and\ \bibinfo {author} {\bibfnamefont {E.}~\bibnamefont
  {Santopinto}},\ }\href {\doibase 10.1103/PhysRevD.90.094022} {\bibfield
  {journal} {\bibinfo  {journal} {Phys. Rev.}\ }\textbf {\bibinfo {volume}
  {D90}},\ \bibinfo {pages} {094022} (\bibinfo {year} {2014})},\ \Eprint
  {http://arxiv.org/abs/1306.2874} {arXiv:1306.2874 [hep-ph]} \BibitemShut
  {NoStop}%
\bibitem [{\citenamefont {Ferretti}\ \emph {et~al.}(2014)\citenamefont
  {Ferretti}, \citenamefont {Galatà},\ and\ \citenamefont
  {Santopinto}}]{Ferretti:2014xqa}%
  \BibitemOpen
  \bibfield  {author} {\bibinfo {author} {\bibfnamefont {J.}~\bibnamefont
  {Ferretti}}, \bibinfo {author} {\bibfnamefont {G.}~\bibnamefont {Galatà}}, \
  and\ \bibinfo {author} {\bibfnamefont {E.}~\bibnamefont {Santopinto}},\
  }\href {\doibase 10.1103/PhysRevD.90.054010} {\bibfield  {journal} {\bibinfo
  {journal} {Phys. Rev.}\ }\textbf {\bibinfo {volume} {D90}},\ \bibinfo {pages}
  {054010} (\bibinfo {year} {2014})},\ \Eprint {http://arxiv.org/abs/1401.4431}
  {arXiv:1401.4431 [nucl-th]} \BibitemShut {NoStop}%
\bibitem [{\citenamefont {Segovia}\ \emph {et~al.}(2015)\citenamefont
  {Segovia}, \citenamefont {Entem},\ and\ \citenamefont
  {Fernández}}]{Segovia:2014mca}%
  \BibitemOpen
  \bibfield  {author} {\bibinfo {author} {\bibfnamefont {J.}~\bibnamefont
  {Segovia}}, \bibinfo {author} {\bibfnamefont {D.~R.}\ \bibnamefont {Entem}},
  \ and\ \bibinfo {author} {\bibfnamefont {F.}~\bibnamefont {Fernández}},\
  }\href {\doibase 10.1103/PhysRevD.91.014002} {\bibfield  {journal} {\bibinfo
  {journal} {Phys. Rev.}\ }\textbf {\bibinfo {volume} {D91}},\ \bibinfo {pages}
  {014002} (\bibinfo {year} {2015})},\ \Eprint {http://arxiv.org/abs/1409.7079}
  {arXiv:1409.7079 [hep-ph]} \BibitemShut {NoStop}%
\bibitem [{\citenamefont {Akbar}\ \emph {et~al.}(2017)\citenamefont {Akbar},
  \citenamefont {Sultan}, \citenamefont {Masud},\ and\ \citenamefont
  {Akram}}]{Akbar:2015evy}%
  \BibitemOpen
  \bibfield  {author} {\bibinfo {author} {\bibfnamefont {N.}~\bibnamefont
  {Akbar}}, \bibinfo {author} {\bibfnamefont {M.~A.}\ \bibnamefont {Sultan}},
  \bibinfo {author} {\bibfnamefont {B.}~\bibnamefont {Masud}}, \ and\ \bibinfo
  {author} {\bibfnamefont {F.}~\bibnamefont {Akram}},\ }\href {\doibase
  10.1103/PhysRevD.95.074018} {\bibfield  {journal} {\bibinfo  {journal} {Phys.
  Rev.}\ }\textbf {\bibinfo {volume} {D95}},\ \bibinfo {pages} {074018}
  (\bibinfo {year} {2017})},\ \Eprint {http://arxiv.org/abs/1511.03632}
  {arXiv:1511.03632 [hep-ph]} \BibitemShut {NoStop}%
\bibitem [{\citenamefont {Deng}\ \emph {et~al.}(2017)\citenamefont {Deng},
  \citenamefont {Liu}, \citenamefont {Gui},\ and\ \citenamefont
  {Zhong}}]{Deng:2016ktl}%
  \BibitemOpen
  \bibfield  {author} {\bibinfo {author} {\bibfnamefont {W.-J.}\ \bibnamefont
  {Deng}}, \bibinfo {author} {\bibfnamefont {H.}~\bibnamefont {Liu}}, \bibinfo
  {author} {\bibfnamefont {L.-C.}\ \bibnamefont {Gui}}, \ and\ \bibinfo
  {author} {\bibfnamefont {X.-H.}\ \bibnamefont {Zhong}},\ }\href {\doibase
  10.1103/PhysRevD.95.074002} {\bibfield  {journal} {\bibinfo  {journal} {Phys.
  Rev.}\ }\textbf {\bibinfo {volume} {D95}},\ \bibinfo {pages} {074002}
  (\bibinfo {year} {2017})},\ \Eprint {http://arxiv.org/abs/1607.04696}
  {arXiv:1607.04696 [hep-ph]} \BibitemShut {NoStop}%
\bibitem [{\citenamefont {Segovia}\ \emph {et~al.}(2016)\citenamefont
  {Segovia}, \citenamefont {Ortega}, \citenamefont {Entem},\ and\ \citenamefont
  {Fernández}}]{Segovia:2016xqb}%
  \BibitemOpen
  \bibfield  {author} {\bibinfo {author} {\bibfnamefont {J.}~\bibnamefont
  {Segovia}}, \bibinfo {author} {\bibfnamefont {P.~G.}\ \bibnamefont {Ortega}},
  \bibinfo {author} {\bibfnamefont {D.~R.}\ \bibnamefont {Entem}}, \ and\
  \bibinfo {author} {\bibfnamefont {F.}~\bibnamefont {Fernández}},\ }\href
  {\doibase 10.1103/PhysRevD.93.074027} {\bibfield  {journal} {\bibinfo
  {journal} {Phys. Rev.}\ }\textbf {\bibinfo {volume} {D93}},\ \bibinfo {pages}
  {074027} (\bibinfo {year} {2016})},\ \Eprint
  {http://arxiv.org/abs/1601.05093} {arXiv:1601.05093 [hep-ph]} \BibitemShut
  {NoStop}%
\bibitem [{\citenamefont {Weng}\ \emph {et~al.}(2019)\citenamefont {Weng},
  \citenamefont {Xiao}, \citenamefont {Deng}, \citenamefont {Chen},\ and\
  \citenamefont {Zhu}}]{Weng:2018ebv}%
  \BibitemOpen
  \bibfield  {author} {\bibinfo {author} {\bibfnamefont {X.-Z.}\ \bibnamefont
  {Weng}}, \bibinfo {author} {\bibfnamefont {L.-Y.}\ \bibnamefont {Xiao}},
  \bibinfo {author} {\bibfnamefont {W.-Z.}\ \bibnamefont {Deng}}, \bibinfo
  {author} {\bibfnamefont {X.-L.}\ \bibnamefont {Chen}}, \ and\ \bibinfo
  {author} {\bibfnamefont {S.-L.}\ \bibnamefont {Zhu}},\ }\href {\doibase
  10.1103/PhysRevD.99.094001} {\bibfield  {journal} {\bibinfo  {journal} {Phys.
  Rev.}\ }\textbf {\bibinfo {volume} {D99}},\ \bibinfo {pages} {094001}
  (\bibinfo {year} {2019})},\ \Eprint {http://arxiv.org/abs/1811.09002}
  {arXiv:1811.09002 [hep-ph]} \BibitemShut {NoStop}%
\bibitem [{\citenamefont {Srivastava}\ \emph {et~al.}(2018)\citenamefont
  {Srivastava}, \citenamefont {Chaturvedi},\ and\ \citenamefont
  {Thakur}}]{Srivastava:2018vxp}%
  \BibitemOpen
  \bibfield  {author} {\bibinfo {author} {\bibfnamefont {P.~K.}\ \bibnamefont
  {Srivastava}}, \bibinfo {author} {\bibfnamefont {O.~S.~K.}\ \bibnamefont
  {Chaturvedi}}, \ and\ \bibinfo {author} {\bibfnamefont {L.}~\bibnamefont
  {Thakur}},\ }\href {\doibase 10.1140/epjc/s10052-018-5912-z} {\bibfield
  {journal} {\bibinfo  {journal} {Eur. Phys. J.}\ }\textbf {\bibinfo {volume}
  {C78}},\ \bibinfo {pages} {440} (\bibinfo {year} {2018})},\ \Eprint
  {http://arxiv.org/abs/1806.09348} {arXiv:1806.09348 [nucl-th]} \BibitemShut
  {NoStop}%
\bibitem [{\citenamefont {Monteiro}\ \emph {et~al.}(2018)\citenamefont
  {Monteiro}, \citenamefont {D'Souza}, \citenamefont {Vidyalakshmi},\ and\
  \citenamefont {Vijaya~Kumar}}]{Monteiro:2018dnx}%
  \BibitemOpen
  \bibfield  {author} {\bibinfo {author} {\bibfnamefont {A.~P.}\ \bibnamefont
  {Monteiro}}, \bibinfo {author} {\bibfnamefont {P.~P.}\ \bibnamefont
  {D'Souza}}, \bibinfo {author} {\bibfnamefont {N.}~\bibnamefont
  {Vidyalakshmi}}, \ and\ \bibinfo {author} {\bibfnamefont {K.~B.}\
  \bibnamefont {Vijaya~Kumar}},\ }\bibfield  {booktitle} {\emph {\bibinfo
  {booktitle} {{Proceedings, 63rd DAE-BRNS Symposium on Nuclear Physics:
  Mumbai, Maharashtra, India, December 10-14, 2018}}},\ }\href@noop {}
  {\bibfield  {journal} {\bibinfo  {journal} {DAE Symp. Nucl. Phys.}\ }\textbf
  {\bibinfo {volume} {63}},\ \bibinfo {pages} {878} (\bibinfo {year}
  {2018})}\BibitemShut {NoStop}%
\bibitem [{\citenamefont {Al-Jamel}(2019)}]{Al-Jamel:2019myn}%
  \BibitemOpen
  \bibfield  {author} {\bibinfo {author} {\bibfnamefont {A.}~\bibnamefont
  {Al-Jamel}},\ }\href {\doibase 10.1142/S0217732319503073} {\bibfield
  {journal} {\bibinfo  {journal} {Mod. Phys. Lett.}\ }\textbf {\bibinfo
  {volume} {2}},\ \bibinfo {pages} {1950307} (\bibinfo {year} {2019})},\
  \Eprint {http://arxiv.org/abs/1908.07707} {arXiv:1908.07707 [hep-ph]}
  \BibitemShut {NoStop}%
\bibitem [{\citenamefont {Chen}\ \emph {et~al.}(2019)\citenamefont {Chen},
  \citenamefont {Zhang},\ and\ \citenamefont {He}}]{Chen:2019uzm}%
  \BibitemOpen
  \bibfield  {author} {\bibinfo {author} {\bibfnamefont {B.}~\bibnamefont
  {Chen}}, \bibinfo {author} {\bibfnamefont {A.}~\bibnamefont {Zhang}}, \ and\
  \bibinfo {author} {\bibfnamefont {J.}~\bibnamefont {He}},\ }\href@noop {} {\
  (\bibinfo {year} {2019})},\ \Eprint {http://arxiv.org/abs/1910.06065}
  {arXiv:1910.06065 [hep-ph]} \BibitemShut {NoStop}%
\bibitem [{\citenamefont {Molina}\ \emph {et~al.}(2017)\citenamefont {Molina},
  \citenamefont {De~Sanctis},\ and\ \citenamefont
  {Fern\'andez-Ram\'irez}}]{molina:2017iaa}%
  \BibitemOpen
  \bibfield  {author} {\bibinfo {author} {\bibfnamefont {D.}~\bibnamefont
  {Molina}}, \bibinfo {author} {\bibfnamefont {M.}~\bibnamefont {De~Sanctis}},
  \ and\ \bibinfo {author} {\bibfnamefont {C.}~\bibnamefont
  {Fern\'andez-Ram\'irez}},\ }\href {\doibase 10.1103/PhysRevD.95.094021}
  {\bibfield  {journal} {\bibinfo  {journal} {Phys. Rev.}\ }\textbf {\bibinfo
  {volume} {D95}},\ \bibinfo {pages} {094021} (\bibinfo {year} {2017})},\
  \Eprint {http://arxiv.org/abs/1703.08097} {arXiv:1703.08097 [hep-ph]}
  \BibitemShut {NoStop}%
\bibitem [{\citenamefont {Dobbs}\ \emph {et~al.}(2012)\citenamefont {Dobbs},
  \citenamefont {Metreveli}, \citenamefont {Seth}, \citenamefont {Tomaradze},\
  and\ \citenamefont {Xiao}}]{Dobbs:2012zn}%
  \BibitemOpen
  \bibfield  {author} {\bibinfo {author} {\bibfnamefont {S.}~\bibnamefont
  {Dobbs}}, \bibinfo {author} {\bibfnamefont {Z.}~\bibnamefont {Metreveli}},
  \bibinfo {author} {\bibfnamefont {K.~K.}\ \bibnamefont {Seth}}, \bibinfo
  {author} {\bibfnamefont {A.}~\bibnamefont {Tomaradze}}, \ and\ \bibinfo
  {author} {\bibfnamefont {T.}~\bibnamefont {Xiao}},\ }\href {\doibase
  10.1103/PhysRevLett.109.082001} {\bibfield  {journal} {\bibinfo  {journal}
  {Phys. Rev. Lett.}\ }\textbf {\bibinfo {volume} {109}},\ \bibinfo {pages}
  {082001} (\bibinfo {year} {2012})},\ \Eprint {http://arxiv.org/abs/1204.4205}
  {arXiv:1204.4205 [hep-ex]} \BibitemShut {NoStop}%
\bibitem [{\citenamefont {Mizuk}\ \emph {et~al.}(2012)\citenamefont {Mizuk}
  \emph {et~al.}}]{Mizuk:2012pb}%
  \BibitemOpen
  \bibfield  {author} {\bibinfo {author} {\bibfnamefont {R.}~\bibnamefont
  {Mizuk}} \emph {et~al.} (\bibinfo {collaboration} {Belle}),\ }\href {\doibase
  10.1103/PhysRevLett.109.232002} {\bibfield  {journal} {\bibinfo  {journal}
  {Phys. Rev. Lett.}\ }\textbf {\bibinfo {volume} {109}},\ \bibinfo {pages}
  {232002} (\bibinfo {year} {2012})},\ \Eprint {http://arxiv.org/abs/1205.6351}
  {arXiv:1205.6351 [hep-ex]} \BibitemShut {NoStop}%
\bibitem [{\citenamefont {Li}\ and\ \citenamefont {Chao}(2009)}]{Li:2009nr}%
  \BibitemOpen
  \bibfield  {author} {\bibinfo {author} {\bibfnamefont {B.-Q.}\ \bibnamefont
  {Li}}\ and\ \bibinfo {author} {\bibfnamefont {K.-T.}\ \bibnamefont {Chao}},\
  }\href {\doibase 10.1088/0253-6102/52/4/20} {\bibfield  {journal} {\bibinfo
  {journal} {Commun. Theor. Phys.}\ }\textbf {\bibinfo {volume} {52}},\
  \bibinfo {pages} {653} (\bibinfo {year} {2009})},\ \Eprint
  {http://arxiv.org/abs/0909.1369} {arXiv:0909.1369 [hep-ph]} \BibitemShut
  {NoStop}%
\bibitem [{\citenamefont {Ferretti}\ \emph {et~al.}(2012)\citenamefont
  {Ferretti}, \citenamefont {Galata}, \citenamefont {Santopinto},\ and\
  \citenamefont {Vassallo}}]{Ferretti:2012zz}%
  \BibitemOpen
  \bibfield  {author} {\bibinfo {author} {\bibfnamefont {J.}~\bibnamefont
  {Ferretti}}, \bibinfo {author} {\bibfnamefont {G.}~\bibnamefont {Galata}},
  \bibinfo {author} {\bibfnamefont {E.}~\bibnamefont {Santopinto}}, \ and\
  \bibinfo {author} {\bibfnamefont {A.}~\bibnamefont {Vassallo}},\ }\href
  {\doibase 10.1103/PhysRevC.86.015204} {\bibfield  {journal} {\bibinfo
  {journal} {Phys. Rev.}\ }\textbf {\bibinfo {volume} {C86}},\ \bibinfo {pages}
  {015204} (\bibinfo {year} {2012})}\BibitemShut {NoStop}%
\bibitem [{\citenamefont {De~Sanctis}\ and\ \citenamefont
  {Quintero}(2010)}]{DeSanctis:2010zz}%
  \BibitemOpen
  \bibfield  {author} {\bibinfo {author} {\bibfnamefont {M.}~\bibnamefont
  {De~Sanctis}}\ and\ \bibinfo {author} {\bibfnamefont {P.}~\bibnamefont
  {Quintero}},\ }\href {\doibase 10.1140/epja/i2010-11032-y} {\bibfield
  {journal} {\bibinfo  {journal} {Eur. Phys. J.}\ }\textbf {\bibinfo {volume}
  {A46}},\ \bibinfo {pages} {213} (\bibinfo {year} {2010})}\BibitemShut
  {NoStop}%
\bibitem [{\citenamefont {Lees}\ \emph {et~al.}(2011)\citenamefont {Lees} \emph
  {et~al.}}]{Lees:2011mx}%
  \BibitemOpen
  \bibfield  {author} {\bibinfo {author} {\bibfnamefont {J.~P.}\ \bibnamefont
  {Lees}} \emph {et~al.} (\bibinfo {collaboration} {BaBar}),\ }\href {\doibase
  10.1103/PhysRevD.84.099901, 10.1103/PhysRevD.84.072002} {\bibfield  {journal}
  {\bibinfo  {journal} {Phys. Rev.}\ }\textbf {\bibinfo {volume} {D84}},\
  \bibinfo {pages} {072002} (\bibinfo {year} {2011})},\ \Eprint
  {http://arxiv.org/abs/1104.5254} {arXiv:1104.5254 [hep-ex]} \BibitemShut
  {NoStop}%
\bibitem [{\citenamefont {Efron}\ and\ \citenamefont
  {Tibshirani}(1994)}]{EfroTibs93}%
  \BibitemOpen
  \bibfield  {author} {\bibinfo {author} {\bibfnamefont {B.}~\bibnamefont
  {Efron}}\ and\ \bibinfo {author} {\bibfnamefont {R.}~\bibnamefont
  {Tibshirani}},\ }\href
  {https://www.crcpress.com/An-Introduction-to-the-Bootstrap/Efron-Tibshirani/p/book/9780412042317}
  {\emph {\bibinfo {title} {An Introduction to the Bootstrap}}},\ Chapman \&
  Hall/CRC Monographs on Statistics \& Applied Probability\ (\bibinfo
  {publisher} {Taylor \& Francis},\ \bibinfo {year} {1994})\BibitemShut
  {NoStop}%
\bibitem [{\citenamefont {Kass}\ \emph {et~al.}(2014)\citenamefont {Kass},
  \citenamefont {Eden},\ and\ \citenamefont {Brown}}]{Kass2014}%
  \BibitemOpen
  \bibfield  {author} {\bibinfo {author} {\bibfnamefont {R.~E.}\ \bibnamefont
  {Kass}}, \bibinfo {author} {\bibfnamefont {U.~T.}\ \bibnamefont {Eden}}, \
  and\ \bibinfo {author} {\bibfnamefont {E.~N.}\ \bibnamefont {Brown}},\
  }\enquote {\bibinfo {title} {Propagation of uncertainty and the bootstrap},}\
  in\ \href {\doibase 10.1007/978-1-4614-9602-1_9} {\emph {\bibinfo {booktitle}
  {Analysis of Neural Data}}}\ (\bibinfo  {publisher} {Springer New York},\
  \bibinfo {address} {New York, NY},\ \bibinfo {year} {2014})\ pp.\ \bibinfo
  {pages} {221--246}\BibitemShut {NoStop}%
\bibitem [{\citenamefont {Fern\'andez-Ram\'irez}\ \emph
  {et~al.}(2016)\citenamefont {Fern\'andez-Ram\'irez}, \citenamefont
  {Danilkin}, \citenamefont {Mathieu},\ and\ \citenamefont
  {Szczepaniak}}]{Fernandez-Ramirez:2015fbq}%
  \BibitemOpen
  \bibfield  {author} {\bibinfo {author} {\bibfnamefont {C.}~\bibnamefont
  {Fern\'andez-Ram\'irez}}, \bibinfo {author} {\bibfnamefont {I.~V.}\
  \bibnamefont {Danilkin}}, \bibinfo {author} {\bibfnamefont {V.}~\bibnamefont
  {Mathieu}}, \ and\ \bibinfo {author} {\bibfnamefont {A.~P.}\ \bibnamefont
  {Szczepaniak}},\ }\href {\doibase 10.1103/PhysRevD.93.074015} {\bibfield
  {journal} {\bibinfo  {journal} {Phys. Rev.}\ }\textbf {\bibinfo {volume}
  {D93}},\ \bibinfo {pages} {074015} (\bibinfo {year} {2016})},\ \Eprint
  {http://arxiv.org/abs/1512.03136} {arXiv:1512.03136 [hep-ph]} \BibitemShut
  {NoStop}%
\bibitem [{\citenamefont {Landay}\ \emph {et~al.}(2017)\citenamefont {Landay},
  \citenamefont {D{\"o}ring}, \citenamefont {Fern\'andez-Ram\'irez},
  \citenamefont {Hu},\ and\ \citenamefont {Molina}}]{Landay:2016cjw}%
  \BibitemOpen
  \bibfield  {author} {\bibinfo {author} {\bibfnamefont {J.}~\bibnamefont
  {Landay}}, \bibinfo {author} {\bibfnamefont {M.}~\bibnamefont {D{\"o}ring}},
  \bibinfo {author} {\bibfnamefont {C.}~\bibnamefont {Fern\'andez-Ram\'irez}},
  \bibinfo {author} {\bibfnamefont {B.}~\bibnamefont {Hu}}, \ and\ \bibinfo
  {author} {\bibfnamefont {R.}~\bibnamefont {Molina}},\ }\href {\doibase
  10.1103/PhysRevC.95.015203} {\bibfield  {journal} {\bibinfo  {journal} {Phys.
  Rev.}\ }\textbf {\bibinfo {volume} {C95}},\ \bibinfo {pages} {015203}
  (\bibinfo {year} {2017})},\ \Eprint {http://arxiv.org/abs/1610.07547}
  {arXiv:1610.07547 [nucl-th]} \BibitemShut {NoStop}%
\bibitem [{\citenamefont {James}\ and\ \citenamefont
  {Roos}(1975)}]{James:1975dr}%
  \BibitemOpen
  \bibfield  {author} {\bibinfo {author} {\bibfnamefont {F.}~\bibnamefont
  {James}}\ and\ \bibinfo {author} {\bibfnamefont {M.}~\bibnamefont {Roos}},\
  }\href {\doibase 10.1016/0010-4655(75)90039-9} {\bibfield  {journal}
  {\bibinfo  {journal} {Comput. Phys. Commun.}\ }\textbf {\bibinfo {volume}
  {10}},\ \bibinfo {pages} {343} (\bibinfo {year} {1975})}\BibitemShut
  {NoStop}%
\bibitem [{\citenamefont {Aaij}\ \emph {et~al.}(2014)\citenamefont {Aaij} \emph
  {et~al.}}]{Aaij:2014hla}%
  \BibitemOpen
  \bibfield  {author} {\bibinfo {author} {\bibfnamefont {R.}~\bibnamefont
  {Aaij}} \emph {et~al.} (\bibinfo {collaboration} {LHCb}),\ }\href {\doibase
  10.1007/JHEP10(2014)088} {\bibfield  {journal} {\bibinfo  {journal} {JHEP}\
  }\textbf {\bibinfo {volume} {10}},\ \bibinfo {pages} {088} (\bibinfo {year}
  {2014})},\ \Eprint {http://arxiv.org/abs/1409.1408} {arXiv:1409.1408
  [hep-ex]} \BibitemShut {NoStop}%
\bibitem [{\citenamefont {Besson}\ \emph {et~al.}(1985)\citenamefont {Besson}
  \emph {et~al.}}]{Besson:1984bd}%
  \BibitemOpen
  \bibfield  {author} {\bibinfo {author} {\bibfnamefont {D.}~\bibnamefont
  {Besson}} \emph {et~al.} (\bibinfo {collaboration} {CLEO}),\ }\href {\doibase
  10.1103/PhysRevLett.54.381} {\bibfield  {journal} {\bibinfo  {journal} {Phys.
  Rev. Lett.}\ }\textbf {\bibinfo {volume} {54}},\ \bibinfo {pages} {381}
  (\bibinfo {year} {1985})}\BibitemShut {NoStop}%
\bibitem [{\citenamefont {Lovelock}\ \emph {et~al.}(1985)\citenamefont
  {Lovelock} \emph {et~al.}}]{Lovelock:1985nb}%
  \BibitemOpen
  \bibfield  {author} {\bibinfo {author} {\bibfnamefont {D.~M.~J.}\
  \bibnamefont {Lovelock}} \emph {et~al.},\ }\href {\doibase
  10.1103/PhysRevLett.54.377} {\bibfield  {journal} {\bibinfo  {journal} {Phys.
  Rev. Lett.}\ }\textbf {\bibinfo {volume} {54}},\ \bibinfo {pages} {377}
  (\bibinfo {year} {1985})}\BibitemShut {NoStop}%
\bibitem [{\citenamefont {Santel}\ \emph {et~al.}(2016)\citenamefont {Santel}
  \emph {et~al.}}]{Santel:2015qga}%
  \BibitemOpen
  \bibfield  {author} {\bibinfo {author} {\bibfnamefont {D.}~\bibnamefont
  {Santel}} \emph {et~al.} (\bibinfo {collaboration} {Belle}),\ }\href
  {\doibase 10.1103/PhysRevD.93.011101} {\bibfield  {journal} {\bibinfo
  {journal} {Phys. Rev.}\ }\textbf {\bibinfo {volume} {D93}},\ \bibinfo {pages}
  {011101} (\bibinfo {year} {2016})},\ \Eprint
  {http://arxiv.org/abs/1501.01137} {arXiv:1501.01137 [hep-ex]} \BibitemShut
  {NoStop}%
\bibitem [{\citenamefont {Chen}\ \emph {et~al.}(2008)\citenamefont {Chen} \emph
  {et~al.}}]{Abe:2007tk}%
  \BibitemOpen
  \bibfield  {author} {\bibinfo {author} {\bibfnamefont {K.~F.}\ \bibnamefont
  {Chen}} \emph {et~al.} (\bibinfo {collaboration} {Belle}),\ }\href {\doibase
  10.1103/PhysRevLett.100.112001} {\bibfield  {journal} {\bibinfo  {journal}
  {Phys. Rev. Lett.}\ }\textbf {\bibinfo {volume} {100}},\ \bibinfo {pages}
  {112001} (\bibinfo {year} {2008})},\ \Eprint {http://arxiv.org/abs/0710.2577}
  {arXiv:0710.2577 [hep-ex]} \BibitemShut {NoStop}%
\bibitem [{\citenamefont {Bruschini}\ and\ \citenamefont
  {González}(2019)}]{Bruschini:2018lse}%
  \BibitemOpen
  \bibfield  {author} {\bibinfo {author} {\bibfnamefont {R.}~\bibnamefont
  {Bruschini}}\ and\ \bibinfo {author} {\bibfnamefont {P.}~\bibnamefont
  {González}},\ }\href {\doibase 10.1016/j.physletb.2019.03.017} {\bibfield
  {journal} {\bibinfo  {journal} {Phys. Lett.}\ }\textbf {\bibinfo {volume}
  {B791}},\ \bibinfo {pages} {409} (\bibinfo {year} {2019})},\ \Eprint
  {http://arxiv.org/abs/1811.08236} {arXiv:1811.08236 [hep-ph]} \BibitemShut
  {NoStop}%
\bibitem [{\citenamefont {Dib}\ and\ \citenamefont {Neill}(2012)}]{Dib:2012vw}%
  \BibitemOpen
  \bibfield  {author} {\bibinfo {author} {\bibfnamefont {C.~O.}\ \bibnamefont
  {Dib}}\ and\ \bibinfo {author} {\bibfnamefont {N.}~\bibnamefont {Neill}},\
  }\href {\doibase 10.1103/PhysRevD.86.094011} {\bibfield  {journal} {\bibinfo
  {journal} {Phys. Rev.}\ }\textbf {\bibinfo {volume} {D86}},\ \bibinfo {pages}
  {094011} (\bibinfo {year} {2012})},\ \Eprint {http://arxiv.org/abs/1208.2186}
  {arXiv:1208.2186 [hep-ph]} \BibitemShut {NoStop}%
\bibitem [{\citenamefont {Anwar}\ \emph {et~al.}(2019)\citenamefont {Anwar},
  \citenamefont {Lu},\ and\ \citenamefont {Zou}}]{Anwar:2018yqm}%
  \BibitemOpen
  \bibfield  {author} {\bibinfo {author} {\bibfnamefont {M.~N.}\ \bibnamefont
  {Anwar}}, \bibinfo {author} {\bibfnamefont {Y.}~\bibnamefont {Lu}}, \ and\
  \bibinfo {author} {\bibfnamefont {B.-S.}\ \bibnamefont {Zou}},\ }\href
  {\doibase 10.1103/PhysRevD.99.094005} {\bibfield  {journal} {\bibinfo
  {journal} {Phys. Rev.}\ }\textbf {\bibinfo {volume} {D99}},\ \bibinfo {pages}
  {094005} (\bibinfo {year} {2019})},\ \Eprint
  {http://arxiv.org/abs/1806.01155} {arXiv:1806.01155 [hep-ph]} \BibitemShut
  {NoStop}%
\bibitem [{\citenamefont {Karliner}\ and\ \citenamefont
  {Rosner}(2015)}]{Karliner:2014lta}%
  \BibitemOpen
  \bibfield  {author} {\bibinfo {author} {\bibfnamefont {M.}~\bibnamefont
  {Karliner}}\ and\ \bibinfo {author} {\bibfnamefont {J.~L.}\ \bibnamefont
  {Rosner}},\ }\href {\doibase 10.1103/PhysRevD.91.014014} {\bibfield
  {journal} {\bibinfo  {journal} {Phys. Rev.}\ }\textbf {\bibinfo {volume}
  {D91}},\ \bibinfo {pages} {014014} (\bibinfo {year} {2015})},\ \Eprint
  {http://arxiv.org/abs/1410.7729} {arXiv:1410.7729 [hep-ph]} \BibitemShut
  {NoStop}%
\bibitem [{\citenamefont {Zhou}\ \emph {et~al.}(2019)\citenamefont {Zhou},
  \citenamefont {Chen},\ and\ \citenamefont {Xiao}}]{Zhou:2018hlv}%
  \BibitemOpen
  \bibfield  {author} {\bibinfo {author} {\bibfnamefont {Z.-Y.}\ \bibnamefont
  {Zhou}}, \bibinfo {author} {\bibfnamefont {D.-Y.}\ \bibnamefont {Chen}}, \
  and\ \bibinfo {author} {\bibfnamefont {Z.}~\bibnamefont {Xiao}},\ }\href
  {\doibase 10.1103/PhysRevD.99.034005} {\bibfield  {journal} {\bibinfo
  {journal} {Phys. Rev.}\ }\textbf {\bibinfo {volume} {D99}},\ \bibinfo {pages}
  {034005} (\bibinfo {year} {2019})},\ \Eprint
  {http://arxiv.org/abs/1810.03452} {arXiv:1810.03452 [hep-ph]} \BibitemShut
  {NoStop}%
\bibitem [{\citenamefont {Godfrey}\ and\ \citenamefont
  {Moats}(2015)}]{Godfrey:2015dia}%
  \BibitemOpen
  \bibfield  {author} {\bibinfo {author} {\bibfnamefont {S.}~\bibnamefont
  {Godfrey}}\ and\ \bibinfo {author} {\bibfnamefont {K.}~\bibnamefont
  {Moats}},\ }\href {\doibase 10.1103/PhysRevD.92.054034} {\bibfield  {journal}
  {\bibinfo  {journal} {Phys. Rev.}\ }\textbf {\bibinfo {volume} {D92}},\
  \bibinfo {pages} {054034} (\bibinfo {year} {2015})},\ \Eprint
  {http://arxiv.org/abs/1507.00024} {arXiv:1507.00024 [hep-ph]} \BibitemShut
  {NoStop}%
\bibitem [{\citenamefont {Hu}\ \emph {et~al.}(2017)\citenamefont {Hu},
  \citenamefont {Leonardo}, \citenamefont {Liu},\ and\ \citenamefont
  {Haytmyradov}}]{Hu:2017pat}%
  \BibitemOpen
  \bibfield  {author} {\bibinfo {author} {\bibfnamefont {Z.}~\bibnamefont
  {Hu}}, \bibinfo {author} {\bibfnamefont {N.~T.}\ \bibnamefont {Leonardo}},
  \bibinfo {author} {\bibfnamefont {T.}~\bibnamefont {Liu}}, \ and\ \bibinfo
  {author} {\bibfnamefont {M.}~\bibnamefont {Haytmyradov}},\ }\href {\doibase
  10.1142/S0217751X17300150} {\bibfield  {journal} {\bibinfo  {journal} {Int.
  J. Mod. Phys.}\ }\textbf {\bibinfo {volume} {A32}},\ \bibinfo {pages}
  {1730015} (\bibinfo {year} {2017})},\ \Eprint
  {http://arxiv.org/abs/1708.02913} {arXiv:1708.02913 [hep-ex]} \BibitemShut
  {NoStop}%
\bibitem [{\citenamefont {Fulsom}(2017)}]{Fulsom:2017erj}%
  \BibitemOpen
  \bibfield  {author} {\bibinfo {author} {\bibfnamefont {B.}~\bibnamefont
  {Fulsom}},\ }in\ \href@noop {} {\emph {\bibinfo {booktitle} {{Proceedings,
  Meeting of the APS Division of Particles and Fields (DPF 2017): Fermilab,
  Batavia, Illinois, USA, July 31 - August 4, 2017}}}}\ (\bibinfo {year}
  {2017})\ \Eprint {http://arxiv.org/abs/1710.00120} {arXiv:1710.00120
  [hep-ex]} \BibitemShut {NoStop}%
\bibitem [{\citenamefont {Ye}(2016)}]{Ye:2016pgb}%
  \BibitemOpen
  \bibfield  {author} {\bibinfo {author} {\bibfnamefont {H.}~\bibnamefont
  {Ye}},\ }\bibfield  {booktitle} {\emph {\bibinfo {booktitle} {{Proceedings,
  24th International Workshop on Deep-Inelastic Scattering and Related Subjects
  (DIS 2016): Hamburg, Germany, April 11-15, 2016}}},\ }\href {\doibase
  10.22323/1.265.0262} {\bibfield  {journal} {\bibinfo  {journal} {PoS}\
  }\textbf {\bibinfo {volume} {DIS2016}},\ \bibinfo {pages} {262} (\bibinfo
  {year} {2016})},\ \Eprint {http://arxiv.org/abs/1607.01740} {arXiv:1607.01740
  [hep-ex]} \BibitemShut {NoStop}%
\bibitem [{\citenamefont {Altmannshofer}\ \emph {et~al.}(2018)\citenamefont
  {Altmannshofer} \emph {et~al.}}]{Kou:2018nap}%
  \BibitemOpen
  \bibfield  {author} {\bibinfo {author} {\bibfnamefont {W.}~\bibnamefont
  {Altmannshofer}} \emph {et~al.} (\bibinfo {collaboration} {Belle II}),\
  }\href@noop {} {\emph {\bibinfo {title} {{The Belle II Physics Book}}}},\
  edited by\ \bibinfo {editor} {\bibfnamefont {E.}~\bibnamefont {Kou}}\ and\
  \bibinfo {editor} {\bibfnamefont {P.}~\bibnamefont {Urquijo}}\ (\bibinfo
  {year} {2018})\ \Eprint {http://arxiv.org/abs/1808.10567} {arXiv:1808.10567
  [hep-ex]} \BibitemShut {NoStop}%
\bibitem [{\citenamefont {Olsen}(2015)}]{Olsen:2014qna}%
  \BibitemOpen
  \bibfield  {author} {\bibinfo {author} {\bibfnamefont {S.~L.}\ \bibnamefont
  {Olsen}},\ }\href {\doibase 10.1007/S11467-014-0449-6} {\bibfield  {journal}
  {\bibinfo  {journal} {Front. Phys.(Beijing)}\ }\textbf {\bibinfo {volume}
  {10}},\ \bibinfo {pages} {121} (\bibinfo {year} {2015})},\ \Eprint
  {http://arxiv.org/abs/1411.7738} {arXiv:1411.7738 [hep-ex]} \BibitemShut
  {NoStop}%
\end{thebibliography}%


%
\end{document}